\documentclass[usegraphicx,usenatbib,useapjfonts,apj]{emulateapj}

\usepackage{apjfonts}
\usepackage{amsbsy}
\usepackage{graphicx}
\usepackage{dcolumn}

\renewcommand\({\left(}
\renewcommand\){\right)}
\renewcommand\[{\left[}
\renewcommand\]{\right]}

\newcommand{\ra}{\rightarrow}

\def\lsim{\raise 0.4ex\hbox{$<$}\kern -0.8em\lower 0.62
ex\hbox{$\sim$}}

\def\gsim{\raise 0.4ex\hbox{$>$}\kern -0.7em\lower 0.62
ex\hbox{$\sim$}}

\def\lbar{{\hbox{$\lambda$}\kern -0.7em\raise 0.6ex
\hbox{$-$}}}

\newcommand\eq[1]{eq.~(\ref{#1})}
\newcommand\eqs[2]{eqs.~(\ref{#1}) and (\ref{#2})}
\newcommand\Eq[1]{Equation~(\ref{#1})}
\newcommand\Eqs[2]{Equations~(\ref{#1}) and (\ref{#2})}

\newcommand\eqss[3]{eqs.~(\ref{#1}), (\ref{#2}) and (\ref{#3})}

\newcommand\pa{\partial}
\newcommand\p{\partial}

\newcommand\ee{\end{equation}}
\newcommand\be{\begin{equation}}
\def\bea{\begin{array}}
\def\eea{\end{array}}\def\ea{\end{array}}
\newcommand\ees{\end{eqnarray}}
\newcommand\bees{\begin{eqnarray}}

\def\p1{{\bf p}_1}
\def\p2{{\bf p}_2}
\def\k1{{\bf k}_1}
\def\k2{{\bf k}_2}

\newcommand{\dddM}{\kern 0.2em \raise 1.9ex\hbox{$...$}\kern -1.0em \hbox{$M$}}
\newcommand{\dddQ}{\kern 0.2em \raise 1.9ex\hbox{$...$}\kern -1.0em \hbox{$Q$}}
\newcommand{\dddI}{\kern 0.2em \raise 1.9ex\hbox{$...$}\kern -1.0em\hbox{$I$}}
\newcommand{\dddJ}{\kern 0.2em \raise 1.9ex\hbox{$...$}\kern-1.0em
\hbox{$J$}}
\newcommand{\dddcalJ}{\kern 0.2em \raise 1.9ex\hbox{$...$}\kern-1.0em
\hbox{${\cal J}$}}

\newcommand{\dddO}{\kern 0.2em \raise 1.9ex\hbox{$...$}\kern -1.0em
\hbox{${\cal O}$}}
\def\dddz{\raise 1.5ex\hbox{$...$}\kern -0.8em \hbox{$z$}}
\def\dddd{\raise 1.8ex\hbox{$...$}\kern -0.8em \hbox{$d$}}
\def\dddbd{\raise 1.8ex\hbox{$...$}\kern -0.8em \hbox{${\bf d}$}}
\def\ddbd{\raise 1.8ex\hbox{$..$}\kern -0.8em \hbox{${\bf d}$}}
\def\dddx{\raise 1.6ex\hbox{$...$}\kern -0.8em \hbox{$x$}}

\def\D{\Delta}
\def\p{\partial}
\def\a{\alpha}

\def\nn{\nonumber}

\def\s{\sigma}

\def\G{\Gamma}
\def\d{\delta}

\def\eps{\epsilon}

\def\dslash{\hspace{-1mm}\not{\hbox{\kern-2pt $\partial$}}}
\def\Dslash{\not{\hbox{\kern-4pt $D$}}}
\def\pslash{\not{\hbox{\kern-2.1pt $p$}}}
\def\kslash{\not{\hbox{\kern-2.3pt $k$}}}
\def\qslash{\not{\hbox{\kern-2.3pt $q$}}}

\newcommand{\inT}{\int_{-\infty}^{\infty}}

\newcommand{\Dl}{\int{\cal D}\lambda}

\newcommand{\fnl}{f_{\rm NL}}

\begin{document}

\title{The Halo Mass Function from  Excursion 
Set Theory.\\
III. Non-Gaussian Fluctuations}
\author{
Michele Maggiore\altaffilmark{1} and
Antonio Riotto\altaffilmark{2,3}
}
\altaffiltext{1}{D\'epartement de Physique Th\'eorique, 
Universit\'e de Gen\`eve, 24 quai Ansermet, CH-1211 Gen\`eve, Switzerland}
\altaffiltext{2}{CERN, PH-TH Division, CH-1211, Gen\`eve 23,  Switzerland}
\altaffiltext{3}{INFN, Sezione di Padova, Via Marzolo 8,
I-35131 Padua, Italy}


\begin{abstract}
We compute  the effect of primordial non-Gaussianity on the 
halo mass function, using 
excursion set theory. In
the presence of non-Gaussianity  the stochastic evolution of the smoothed density field, as a function of  the smoothing scale,  is
non-markovian and beside ``local'' terms that generalize Press-Schechter (PS) theory,
there are also   ``memory'' terms, whose effect on the mass function can be  computed using the  formalism developed in the first paper of this series. 
We find that, when computing the effect of the three-point correlator on the mass function, a PS-like approach which consists in neglecting the cloud-in-cloud problem and in multiplying the final result by a fudge factor $\simeq 2$, is in principle not justified. When computed correctly in the framework of excursion set theory, in fact, the ``local" contribution vanishes (for all odd-point  correlators
the contribution of the image gaussian cancels the Press-Schechter contribution rather than adding up), and the result
comes entirely from  non-trivial memory terms which are absent in PS theory. However it turns out that, in the limit of large halo masses, where the effect of non-Gaussianity is more relevant,
these memory terms give a contribution  which is the the same as that computed naively with PS theory, plus subleading terms depending on derivatives of the three-point correlator. 
We finally combine these results with 
the diffusive barrier model developed in the second paper of this
series, and we find that
the resulting mass function reproduces recent  $N$-body simulations 
with non-Gaussian initial conditions,
without the introduction of any ad hoc 
parameter.
\end{abstract}



\section{Introduction}

In the first two papers  of this series
(\cite{MR1} and \cite{MR2}, papers I and II in the following) we have
studied the  mass function of
dark matter halos using the excursion set formalism.
The halo mass function
can be written as
\be\label{dndMdef}
\frac{dn(M)}{dM} = f(\s) \frac{\bar{\rho}}{M^2} 
\frac{d\ln\s^{-1} (M)}{d\ln M}\, ,
\ee
where $n(M)$ is the  number density of dark matter halos of mass $M$,
$\s(M)$ is the variance of the linear density field smoothed on a
scale $R$ corresponding to a mass $M$, and
$\bar{\rho}$ is the average density of the universe. 
Analytical derivations of the halo mass function 
are typically based  on
Press-Schechter (PS) 
theory \citep{PS} and its extension~\citep{PH90,Bond} known as excursion set
theory  (see 
\cite{Zentner} for a recent review).
In  excursion set theory the density
perturbation evolves stochastically 
with the smoothing scale,
and the problem of computing the probability of halo formation is
mapped into the so-called first-passage time problem in the presence
of a barrier. With this method, for gaussian fluctuations one obtains
\be\label{fps}
f_{\rm PS}(\s) = 
\(\frac{2}{\pi}\)^{1/2}\, 
\frac{\d_c}{\s}\, 
\, e^{-\d_c^2/(2\s^2)}\, ,
\ee
where $\d_c\simeq 1.686$ is the critical value in the spherical collapse
model. This result can be extended to arbitrary redshift
$z$ by reabsorbing the evolution of the variance into $\d_c$, so that
$\d_c$ in the above result is replaced by $\d_c(z)=\d_c(0)/D(z)$, where
$D(z)$ is the linear growth factor.
\Eq{fps}  is only valid when the density contrast 
is smoothed with a sharp filter in momentum space.
In this case the evolution of the density contrast $\d(R)$ with the
smoothing scale is  
markovian, and  the  probability that the density contrast reaches a
given value $\d$ at a given smoothing scale
satisfies a
Fokker-Planck 
equation with an ``absorbing barrier'' boundary condition. From the 
solution of this equation one
obtains \eq{fps}, including a well-known factor of two that
Press and Schecther were forced to add by hand.

However, as is well-known, a sharp filter in momentum space is not
appropriate for comparison with experimental data from upcoming galaxy
surveys, nor with $N$-body
simulations, because it is not possible to associate unambiguously a
mass $M$ to the smoothing scale $R$ used in this filter. Rather, one
should use a tophat filter in coordinate space, in which case the mass
associated to a smoothing scale $R$ is trivially $(4/3)\pi
R^3\rho$. 
If one wants to compute
the halo mass function with a tophat filter in coordinates space 
one is confronted with a much more difficult problem, where the
evolution of $\d$ with the smoothing scale 
is no longer markovian~\citep{Bond}. 
Nevertheless, in paper~I we succeeded in
developing a formalism that allows us to compute  perturbatively
these non-markovian
effects and, for gaussian fluctuations, we found that, to first order,
\eq{fps} is modified to
\be\label{ourfI}
f(\s) = (1-\kappa)
\(\frac{2}{\pi}\)^{1/2}\, 
\frac{\d_c}{\s}\, 
\, e^{-\d_c^2/(2\s^2)}
+\frac{\kappa}{\sqrt{2\pi}}\, 
\frac{\d_c}{\s}\, \G\(0,\frac{\d_c^2}{2\s^2}\)\, ,
\ee
where 
\be\label{adiRlim}
\kappa (R) \equiv
\lim_{R'\ra \infty}
\frac{\langle\d(R')\d(R)\rangle}{\langle\d^2(R')\rangle} -1
\simeq 0.4592-0.0031\, R\, ,
\ee
$R$ is measured in ${\rm Mpc}/h$, $\G(0,z)$ is the
incomplete Gamma function, and the numerical value
of $\kappa(R)$ is computed using
a tophat filter function in coordinate space and
a $\Lambda$CDM model with 
$\s_8=0.8$,
$h=0.7$, $\Omega_M=1-\Omega_{\Lambda}=0.28$, $\Omega_B=0.046$
and $n_s=0.96$,
consistent with the WMAP 5-years data release. 

This analytical result reproduces  well the result of a
Monte Carlo realization of the
first-crossing distribution of excursion set theory,
obtained by integrating numerically a
Langevin equation with a colored noise, performed in
\cite{Bond} and in \cite{RKTZ}. This is a  useful test of
our technique.
Still, neither \eq{fps} nor \eq{ourfI} perform well when compared to
cosmological $N$-body simulation, which means that some crucial physical
ingredient is still missing in the model.
This is not surprising, since the spherical (or ellipsoidal) collapse
model is a very simplified description of the process of halo
formation which, as shown by N-body simulations, 
is much more complicated, and proceeds
through a mixture of smooth accretion and violent encounters leading to
merging as well as to fragmentation 
(see \cite{Springel:2005nw}
and the related movies at 
http://www.mpa-garching.mpg.de/galform/millennium/). 
Furthermore, the very operative definition of what is a dark matter halo is a
subtle issue. Real halos are not spherical. They are at best triaxial,
and often much more irregular, expecially if they experienced recent
mergers. Searching for them with a spherical overdensity algorithm
therefore introduces further statistical uncertainties. Similar
considerations hold for Friends-of-Friends halo finders. 

In paper~II we have discussed in detail these uncertainties and we
have proposed that at least some of the complications intrinsic to a
realistic process of halo formation (as well as the statistical
uncertainties related to the details of the halo finder) 
can be accounted for, within the
excursion set framework, by
treating the critical value
for collapse as a stochastic variable. In this approach all our
ignorance on the details of halo formation is buried into the variance
of the fluctuations of the collapse barrier. 
The computation of the halo mass function is then  mapped 
into a first-passage time process in the presence of a diffusing
barrier,
i.e. a barrier whose
height evolves according to a diffusion equation. For gaussian fluctuations we  found that
\eq{ourfI} must be replaced by
\bees\label{ourf}
f(\s) &=& (1-\tilde{\kappa})
\(\frac{2}{\pi}\)^{1/2}\, 
\frac{a^{1/2}\d_c}{\s}\, 
\, e^{-a\d_c^2/(2\s^2)}\nn\\
&&+\frac{\tilde{\kappa}}{\sqrt{2\pi}}\,
\frac{a^{1/2}\d_c}{\s}\,
 \G\(0,\frac{a \d_c^2}{2\s^2}\)\, ,
\ees
where  
\be\label{aDB}
a=\frac{1}{1+D_B}\, ,\hspace{15mm}
\tilde{\kappa}=\frac{\kappa}{1+D_B}\, ,
\ee
and $D_B$ is an effective diffusion coefficient for the
barrier. A first-principle computation of $D_B$ appears difficult, but
from recent studies of the properties of the collapse barrier in
$N$-body simulations \citep{RKTZ} we deduced a value
$D_B\simeq (0.3\d_c)^2$. 
Using this value
for $D_B$ in \eq{aDB} gives $a\simeq 0.80$, so
\be\label{a079}
\sqrt{a}\simeq 0.89\, .
\ee
We see that the net effect of the diffusing barrier is that, in the mass function, $\d_c$ is replaced by $a^{1/2}\d_c$, which is the replacement that was made by hand, simply to fit the data, in \cite{ST,SMT}. 

The above result was obtained by considering a barrier that fluctuates over the constant value $\d_c$ of the spherical collapse model. More generally, one should consider fluctuations over the barrier $B(\s)$  given by the ellipsoidal collapse model. Since the latter reduces to the former in the small $\s$ limit (i.e. for large halo masses), \eq{ourf} is better seen as the large mass limit of a more accurate mass function obtained from a barrier that flucutates around the average value $B(\s)$ given by the ellipsoidal collapse model. When $\kappa=0$ \eq{ourf} reduces to the large mass limit of the Sheth-Tormen mass function. So, \eq{ourf} generalizes the 
Sheth-Tormen mass function by taking into account the effect of the tophat filter  
 in coordinate space, while \eq{aDB} provides a physical motivation for the introduction of the parameter $a$.

\Eq{ourf} is in excellent agreement with the $N$-body
simulations for gaussian primordial fluctuations, see
Figs.~6 and 7 of paper~II. 
 We stress that our value 
$a\simeq 0.80$ is
not determined by  fitting the mass function  to the data. We do have an input from
the $N$-body simulation here, which is however quite indirect, and is the
measured variance of the threshold for collapse, which for small $\s$
is determined in \cite{RKTZ} to be $\Sigma_B\simeq 0.3\s$. Our
diffusing barrier model of paper~II translates this information into an
effective diffusion coefficient for the barrier,
$D_B=(0.3\d_c)^2$, and predicts $a=1/(1+D_B)$. We refer the reader to
paper~II for details and discussions of the physical motivations for
the introduction of a stochastic barrier.

The above results refer to initial density fluctuations which have a gaussian distribution.
In this paper we attack the problem of the effect on the halo mass
function of  non-Gaussianities in the primordial density field. 
Over the last
decade a great deal of  evidence has been accumulated from the Cosmic Microwave
Background (CMB) anisotropy and Large Scale Structure (LSS) spectra
that the observed
structures originated from seed fluctuations generated during a primordial
stage of inflation. While standard one-single field models of
slow-roll inflation
predict that these fluctuations are very close to  
gaussian (see \cite{acquaviva,maldacena}), 
non-standard scenarios allow for a larger level of non-Gaussianity
(see \cite{bartoloreview} 
and refs. therein).  Deviations from non-Gaussianity are usually parametrized by a dimensionless
quantity $f_{\rm NL}$ (\cite{bartoloreview}) whose value sets the magnitude of the three-point
correlation function. If the process generating the primordial non-Gaussianity  
is local in space, the parameter  $f_{\rm NL}$ in Fourier space is 
independent from the momenta
entering the three-point correlation function; if instead the process is
non-local in space, like in 
models of inflation with non-canonical kinetic terms, $f_{\rm NL}$
acquires a dependence on the momenta. It is clear that  detecting a
significant amount of 
non-Gaussianity and its shape either from the CMB or from the
LSS offers the possibility of opening a   window  into the 
dynamics of the universe
during the very first 
stages of its
evolution. Current
limits on the strength of non-Gaussianity set the $f_{\rm NL}$
parameter to be smaller 
than ${\cal O}(100)$ (\cite{wmap5}).

Non-Gaussianities are particularly relevant in the  high-mass end of
the power spectrum of perturbations, i.e. on the scale of galaxy clusters,
since the effect of non-Gaussian fluctuations becomes especially visible on
the  tail of the probability distribution. 
As a result, both 
the abundance and  the clustering properties of very massive halos
are sensitive probes of primordial 
non-Gaussianities \citep{MLB,GW,LMV,MMLM,KOYAMA,MVJ,RB,RGS}, 
and could be detected or significantly constrained by
the various planned large-scale galaxy surveys,
both ground based (such as DES, PanSTARRS and LSST) and on satellite
(such as EUCLID and ADEPT) see,  e.g.  \cite{Dalal} and \cite{CVM}. 
Furthermore, the primordial non-Gaussianity
alters the clustering of dark matter halos inducing a scale-dependent
bias on large 
scales \citep{Dalal,MV,slosar,tolley}  while even for small primordial
non-Gaussianity the evolution of perturbations on super-Hubble scales yields extra
contributions on 
smaller scales (\cite{bartolosig}).

At present, there exist already various $N$-body simulations where
non-Gaussianity has been included in the initial
conditions \citep{kang,Gros,Dalal,des,PPH,Grossi} and which are useful
to test the accuracy of the
different theoretical predictions for the dark matter halo mass
function with non-Gaussianity.

Various attempts at computing analytically the effect 
of primordial  non-Gaussianities on the mass
function  exist in the literature, based on
non-Gaussian extensions of PS 
theory~\citep{Chiu:1997xb,RB,MVJ,LoVerde}.
However, for  gaussian flucutations, in the large mass regime
PS theory
is off by one order of
magnitude. It is clear 
that, by computing  non-Gaussian
corrections over a theory that, already at the gaussian level, 
in the relevant regime is off by
an order of magnitude, one cannot hope to
get the correct mass function for the
non-Gaussian case. What is typically done in the 
recent literature is to take the
ratio $R_{\rm NG}(M)$
of the non-Gaussian halo mass function to the gaussian 
halo mass function, both computed
within the framework of PS theory, hoping that 
even if neither  the former nor the latter
are correct, still
their ratio might catch  the  main
modifications due to  non-Gaussianities. The full non-Gaussian halo
mass function is then obtained by taking  a fit to
the data in the gaussian case, such as the Sheth and Thormen mass
function \citep{ST,SMT}, and multiplying it by $R_{\rm NG}(M)$.
With this philosophy, the result of
\cite{MVJ} reads\footnote{We thank S. Matarrese for pointing out to us a
  typo in \cite{MVJ}.}
\bees
&&R_{\rm NG}(\s)=\exp\left\{ \frac{\d_{c}^3\, {\cal S}_3(\s)}{6\s^2}
\right\}\label{RNG:MVJ}\\
&&\times
\left| \frac{1}{6}
\frac{\d_{c}}{\sqrt{1-\d_{c}{\cal S}_3(\s)/3}} 
\frac{d{\cal S}_3}{d\ln \sigma}  
+ \sqrt{1-\d_c{\cal S}_3(\s)/3}\right| \, , 
\nonumber 
\ees
where
\be\label{defSk}
{\cal S}_3(\s) =\frac{\langle \d^3(S)\rangle}{\langle \d^2(S)\rangle^2}
\ee
is the (normalized) skewness of the density field and, as usual,
$S=\s^2$ is the  variance. Since $\s=\s(M)$,
we can equivalently consider $R_{\rm NG}$ as a 
function of $M$.\footnote{We  do not write explicitly the
dependence of $R_{\rm NG}(\s)$ on redshift $z$, which enters through
the variance $\s^2$ and, as usual, can be reabsorbed into the height 
$\d_c$ of
the critical value for collapse. The normalized skewness 
must instead be taken at $z=0$, see \cite{Grossi}.}

With a similar philosophy, but a different expansion technique, namely
the Edgeworth expansion, \cite{LoVerde} propose
\be\label{RNG:LoVerde}
R_{\rm NG}(\s)=1+\frac{1}{6}\frac{\s^2}{\delta_{c}} 
\left[{\cal S}_3(\s)
\left(\frac{\delta_{c}^4}{\s^4} 
 -\frac{2\delta_{c}^2}{\s^2} -1\right)+
\frac{d{\cal S}_3}{d \ln \sigma}
\left(\frac{\delta_{c}^2}{\s^2}-1\right)\right]\, . 
\ee
In the limit $\s/\d_c\ll 1$, \eq{RNG:LoVerde} becomes
\be
R_{\rm NG}(\s)=1+\frac{\delta_{c}^3{\cal S}_3(\s)}{6\s^2}\, .
\ee
The same result is obtained from 
\eq{RNG:MVJ} expanding first to linear order in
${\cal S}_3(\s)$, and then retaining the leading term of the expansion
for small  $\s/\d_c$. The two formulas 
differ instead at the level of the terms subleading  in
the expansion for small $\s/\d_c$.
In \cite{Grossi}, in order to fit the data of $N$-body simulations, 
it was suggested to modify both \eq{RNG:MVJ} and \eq{RNG:LoVerde},
by making the replacement
\be\label{dradec}
\d_c\ra \delta_{\rm eff}=\sqrt{a} \,\d_c\, ,
\ee
with a value $\sqrt{a}\simeq 0.86$ obtained from the fit to the
data, very close to our prediction given in
\eq{a079}.\footnote{As discussed in paper~II, \label{foot:a}
the value of the diffusion constant of the barrier $D_B$, and hence our prediction for $a$, depends on the halo finder. The value $D_B\simeq 0.25$, which leads to $\sqrt{a}\simeq 0.89$, has been deduced from the simulation of \cite {RKTZ}, that uses a spherical overdensity (SO) halo finder with $\D=200$, while \cite{Grossi} use a friends-of-friends (FOF) halo finder with link-length 0.2. In the gaussian case, the mass functions obtained from these two finders are very close to each other. However, in order to perform  an accurate numerical comparison of our prediction  with $N$-body simulations
with non-Gaussian initial conditions,  in would be necessary to determine both $D_B$ and the mass function with the same halo finder.}
In the gaussian case  we have shown in paper~II that this
replacement, which in the previous literature was made ad hoc just to
fit the data,  actually follows from  the diffusive barrier model, 
see \eq{ourf}, and that the precise value of $a$ depends, among other
things, on the details of the halo finder in the simulation, so
(slightly) different values of $a$ are obtained from $N$-body simulations
with different halo finders. Below we will see how the results of paper~II
generalize to the non-Gaussian case.

In \cite{Grossi} it is shown that, after performing  the
replacement (\ref{dradec}), both \eq{RNG:MVJ} and \eq{RNG:LoVerde}
are in  good agreement
with the result of $N$-body simulations with non-Gaussian initial
conditions, which a posteriori can be seen as a justification of the
procedure used in their derivation.
However, it is clear that 
taking the ratio of two  results that, in the interesting mass range, are
known to be both off by one order of magnitude, in order to get a fine
effect such as the non-Gaussian corrections,
can only be considered  as a
heuristic procedure. First of all,
PS theory by itself produces a
wrong exponential factor, since it would give $a=1$. Here one might  argue that the gaussian and non-Gaussian
mass functions have the same exponential behavior, so this effect
cancels when considering the ratio $R_{\rm NG}$, and is anyhow
accounted for by the heuristic prescription (\ref{dradec}). Still, a
further source of concern is that the derivation of the PS mass function 
in \cite{Bond} requires that the density field $\d$ evolves
with the smoothing scale $R$ (or more precisely with 
$S(R)$) in a markovian way. Only under this assumption one can derive
\eq{fps} together with the correct factor of two that Press and
Schechter were forced to introduce by hand.
As we have discussed at length in paper~I, this markovian assumption
is broken by the use of a filter function different from
a sharp filter in momentum space and, of course, it is further
violated by the inclusion of non-Gaussian corrections. When studying non-Gaussianities, 
it is therefore particularly important to perform the computation including the effect of the filter, otherwise one would attribute to primordial non-Gaussianities effects on the mass functions which are due, more trivially, to the filter.

The formalism that we have developed in papers~I and ~II, however,
allows us to attack the problem. First of all,
in paper~I we have set up a ``microscopic'' approach which is in
principle exact.\footnote{By exact we mean that, given
the problem of halo formation as it is formulated within
excursion set theory, the path integral technique developed in paper~I  is an exact way of attacking the mathematical problem of the first-passage of a barrier by trajectories performing a non-markovian stochastic motion (at least order by order in the non-markovian corrections).
Of course, one should not forget that excursion set
theory itself gives only an approximate description of the physics involved.}
 With this formalism, we
computed the non-markovian corrections due to the
filter function, which are given by the terms proportional to $\kappa$
in \eq{ourfI}.
This is important because it allows us to subtract, from a measurement
of the halo mass function, the ``trivial'' effect due to the filter,
and to remain with the effects  due
to genuine non-Gaussianities. Second, putting together the corrections
due to the filter with the model of a diffusing barrier, we ended up
with a halo mass function which works very well in the gaussian case,
see Figs.~6 and 7
 of paper~II, and which therefore is a meaningful
starting point for the inclusion of non-Gaussian perturbations.
Finally, the formalism developed in paper~I can be applied,
with simple modifications, to the perturbative computation
of the non-Gaussian corrections. This will be the subject of the
present paper.

This paper is organized as follows. In Section~\ref{sect:path},
extending to the non-Gaussian case  the results presented
in paper~I,  we show
how to formulate the first-passage time
problem for non-Gaussian fluctuations
in terms of a path integral with
boundaries, and we recall the basic points of the computation of
non-markovian corrections performed in paper~I.
In Section~\ref{sect:excNG}
we compute the non-Gaussian corrections with the excursion set method, and we
present our results for the halo mass function. We will see that, in
the approximation in which the three-point correlator at different
times is replaced by the corresponding cumulant, we recover
\eq{RNG:LoVerde} exactly, including the replacement (\ref{dradec}),
except that now this replacement is not performed ad hoc to fit the
$N$-body simulations, but is the consequence of the diffusing barrier
model of paper~II, which also predicts $\sqrt{a}\simeq 0.89$, in
remarkable agreement with the findings of \cite{Grossi}.\footnote{The
parameter that we denote by $a$ is the same as the parameter $q$
of \cite{Grossi}.} We will see however that this result comes out in a rather unexpected way. In fact, the ``local'' term that, in excursion set theory, is supposed to give back the PS result multiplied by the appropriate factor of two, actually vanishes, because 
for the three-point correlator (as well as for all odd-point correlators) the contribution 
of the image gaussian cancels the Press-Schechter contribution rather than adding up. The result (\ref {RNG:LoVerde}) comes 
entirely from non-trivial memory terms, that have no correspondence in the naive PS approach.

We will then go beyond the approximation in which 
the three-point correlator at different
times is replaced by the corresponding cumulant, by
computing explicitly the mass function  at next-to-leading order and at
next-to-next-to-leading order in the small parameter $\s^2/\d_c^2$.
We wil then find further corrections, which depends on the derivative of the correlator, and which,  with respect to the small parameter $\s^2/\d_c^2$, are of the same order as the subleading terms given in \eq {RNG:LoVerde}.

Finally, in Section~\ref{sect:concl} we present our conclusions,
summarizing the findings of this series of three papers.

The focus of this paper is on the generalization of excursion set
theory to non-Gaussian fluctuations. However,
in appendix~A we  examine, with our path
integral formalism, the generalization of naive PS
theory to non-Gaussian fluctuations, and we will contrast it with
the generalization of excursion set
theory.

We have attempted to write this paper in a reasonably self-contained
manner, but the reading of this paper will certainly be facilitated
by a previous acquaintance with the first two papers of this series,
in particular with paper~I.

\section{Path integral approach to stochastic problems.
Non-Gaussian fluctuations}\label{sect:path}

\subsection{General formalism}

In this section we extend to non-Gaussian fluctuations the path
integral approach that we developed in Section~3 of paper~I for
gaussian fluctuations. Our notation is as in paper~I. In particular, we consider the density field $\d$  smoothed over a radius $R$ with a tophat filter in coordinate space. We denote by $S$ the variance of the smoothed density field and, as usual in  excursion set theory, we consider $\d$ as a variable evolving stochastically with respect to the ``pseudotime" $S$, 
(see e.g. Sections~2 of paper~I).
The statistical properties of a random variable $\d(S)$
are specified by its connected 
correlators 
\be\label{corrdifft}
\langle \d (S_1)\ldots\d (S_{p})
\rangle_c\, ,
\ee
where the subscript $c$ stands for ``connected''.
We will also use the notation
\be\label{defmup}
\langle \d^p(S)\rangle_c \equiv \mu_p(S)\, ,
\ee
when all arguments $S_1, S_2, \ldots$ are equal. The quantities
$\mu_p(S)$ are
also called the cumulants. 
As in paper~I, we consider an ensemble of
trajectories all starting at $S_0=0$ from an initial position
$\d(0) =\d_0$ (we will typically choose $\d_0=0$ but the computation can be performed in full generality) 
and we follow them for a ``time'' $S$.  
We discretize the interval $[0,S]$ in steps
$\D S=\eps$, so $S_k=k\eps$ with $k=1,\ldots n$, and $S_n\equiv S$. 
A trajectory is  then defined by
the collection of values $\{\d_1,\ldots ,\d_n\}$, such that $\d(S_k)=\d_k$.

The probability density in the space of  trajectories is 
\be\label{defW}
W(\d_0;\d_1,\ldots ,\d_n;S_n)\equiv \langle
\d_D (\d(S_1)-\d_1)\ldots \d_D (\d(S_n)-\d_n)\rangle\, ,
\ee
where  $\d_D$
denotes the Dirac delta.
As in paper~I, our basic object will be
\be\label{defPi}
\Pi_{\eps} (\d_0;\d_n;S_n)
 \equiv\int_{-\infty}^{\d_c} d\d_1\ldots \int_{-\infty}^{\d_c}d\d_{n-1}\, 
W(\d_0;\d_1,\ldots ,\d_{n-1},\d_n;S_n).
\ee
The usefulness of $\Pi_{\eps}$ is that it allows us to compute the
first-crossing rate from first principles, without the need of
postulating the existence of an absorbing barrier. In fact,
the quantity
\be
\int_{-\infty}^{\d_c}d\d_n\, \Pi_{\eps}(\d_0;\d_n;S_n)
\ee
gives the probability that at ``time'' $S_n$ a trajectory always stayed in
the region $\d<\d_c$, for all times' smaller than $S_n$. The rate of change
of this quantity is therefore equal to minus the rate at which trajectories
cross for the first time the barrier, so 
the first-crossing rate is
\be\label{calFTPi}
{\cal F}(S_n)=-\frac{\pa}{\pa S_n} \int_{-\infty}^{\d_c}d\d_n\, 
\Pi_{\eps}(\d_0;\d_n;S_n)\, .
\ee
The halo mass function is then obtained from the first-crossing rate
using \eq{dndMdef} together with (see eq.~(33) of paper~I)
\be\label{deffs}
f(\s)=2\s^2{\cal F}(\s^2)\, ,
\ee
where $S=\s^2$.
For comparison, it is also useful to introduce
\be\label{defPi0}
\Pi_{{\rm PS},\eps} (\d_0;\d_n;S_n)
 \equiv\inT d\d_1\ldots d\d_{n-1}\, W(\d_0;\d_1,\ldots ,\d_{n-1},\d_n;S_n)\, .
\ee
So, $\Pi_{{\rm PS},\eps}(\d_0;\d_n;S_n)$ is  the probability density  of arriving in
$\d_n$  at time $S_n$,  starting from $\d_0$ at time $S_0=0$,
through any possible trajectory, while 
$\Pi_{\eps} (\d_0;\d_n;S_n)$ is the probability density of arriving in
$\d_n$ at time $S$,  again starting from $\d_0$ at time $S_0=0$, 
through trajectories that never exceeded $\d_c$.
Observe that in both cases the final point
$\d_n$  ranges over $-\infty<\d_n<\infty$. 
Inserting \eq{defW} into \eq{defPi0} and carrying out the integrals
over
$d\d_1\ldots d\d_{n-1}$ we see that
\be
\Pi_{{\rm PS},\eps}(\d_0;\d_n;S_n) = \langle \d_D (\d(S_n)-\d_n)\rangle\, .
\ee
Therefore $\Pi_{{\rm PS},\eps}$
can depend only on
the correlators (\ref{corrdifft}) with all times equal to $S_n$, 
i.e. on the cumulants $\mu_p(S_n)$.
In contrast,
$\Pi_{\eps}(\d_0;\d_n;S_n)$ is a much more complicated object, that
depends on the multi-time correlators given in \eq{corrdifft}.

Furthermore, we see that $\Pi_{{\rm PS},\eps}$ is 
actually independent of $\eps$, since the
integration over the intermediate positions has been carried out
explicitly, and the result depend only on $\d_n$ and $S_n$. Thus, we
will write $\Pi_{{\rm PS},\eps}$ simply as $\Pi_{\rm PS}$. In contrast,
$\Pi_{\eps}$ depends on $\eps$, and we keep this $\eps$ 
dependence explicit.
We are finally interested in its continuum limit, $\Pi_{\eps=0}$, and
we have already seen in paper~I 
that taking the limit $\eps\ra 0$ 
of $\Pi_{\eps}$ is non-trivial.
So, despite their formal similarity, $\Pi_{\eps}$ and $\Pi_{\rm PS}$ are two
very different objects. The distribution function $\Pi_{\rm PS}$ has a
trivial continuum limit, and depend only on the cumulants, while
$\Pi_{\eps}$ depends on the full correlation functions
(\ref{corrdifft}), and its continuum limit is non-trivial. All the
complexity enters in $\Pi_{\eps}$ through the presence of a boundary
in the integration domain, since the variables $\d_i$ are integrated
only up to $\d_c$. 

The use of  $\Pi_{\rm PS}$ generalizes to non-Gaussian
fluctuations  the original PS theory, since we are integrating
over all trajectories, including trajectories that perform
multiple up- and down-crossings of the critical 
value $\d_c$, and therefore suffers of the same cloud-in-cloud problem
of the original PS theory. In the 
literature~\citep{Chiu:1997xb,RB,MVJ,LoVerde} this density functional
has then been  used 
together with the ad hoc prescription that we must multiply the mass
function derived from it by a ``fudge factor'' that ensures that the total
mass of the universe ends up in virialized objects. For gaussian
fluctuations this is the well-known factor of two of Press and
Schechter, while for non-Gaussian theories it is different, although
typically close to two.

In contrast, $\Pi_{\eps}$ generalizes to non-Gaussian fluctuations the
approach of the excursion set method, where the ``cloud-in-cloud''
problem is cured focusing on the first-passage time  of the trajectory,
and no ad hoc multiplicative factor is required. So, 
$\Pi_{\eps}$ is the correct quantity to compute.
From the comparison of $\Pi_{\eps}$ and $\Pi_{\rm PS}$ performed
above, we understand that the difference between the two is not just
a matter of an overall normalization factor.  As we have seen above,
in $\Pi_{\rm PS}$ all the information contained in the correlators at
different ``times'' get lost, since it depends only on the cumulants. The
correlators at different time contain, however, important physical
information. Recalling that the role of ``time'' is actually played by
$S(R)$, the correlators at different time are actually correlators
between density fields at different smoothing scales $R_1$, $R_2$,
etc., and therefore carry  the information on the dependence
of halo formation  on the environment and on the past history.
These informations are intrinsically non-markovian, which is the
reason why 
$\Pi_{\eps}$ is much more difficult to compute. However,
these correlations are
physically very important, especially when we study the
non-Gaussianities, and are completely lost in the extension of PS
theory based on $\Pi_{\rm PS}$. For this reason, our real interest is
in computing  the
distribution function $\Pi_{\eps}$, while 
$\Pi_{\rm PS}$ will only be considered as
a benchmark against which we can compare the  results provided
by $\Pi_{\eps}$.

The first problem that we address is how to express  
$\Pi_{\rm PS} (\d_0;\d;S)$ and  $\Pi_{\eps} (\d_0;\d;S)$,
in terms of
the correlators  of the theory.
Using the integral representation of the Dirac delta
\be
\d_D (x)=\inT\frac{d\lambda}{2\pi}\, e^{-i\lambda x}\, ,
\ee
we write \eq{defW} as
\be\label{compW}
W(\d_0;\d_1,\ldots ,\d_n;S_n)=\inT\frac{d\lambda_1}{2\pi}\ldots\frac{d\lambda_n}{2\pi}\, 
e^{i\sum_{i=1}^n\lambda_i\d_i} \langle e^{-i\sum_{i=1}^n\lambda_i\d(S_i)}\rangle
\, .
\ee
We must therefore compute
\be
e^Z\equiv \langle e^{-i\sum_{i=1}^n\lambda_i\d(S_i)}\rangle \, .
\ee
This is a well-known object both
in quantum field theory and in statistical mechanics, 
since it is  the generating
functional of the connected Green's functions, see e.g.
\cite{Stratonovich}. To a field theorist
this is even more clear
if we define the ``current'' $J$ from $-i\lambda =\eps J$, and we
use a continuous
notation, so that
\be
e^Z =\langle e^{i\int dS\, J(S)\d(S)}\rangle \, .
\ee 
Therefore
\bees
Z&=&\sum_{p=2}^{\infty} \frac{(-i)^p}{p!}\, 
 \sum_{i_1=1}^n\ldots \sum_{i_p=1}^n
\lambda_{i_1}\ldots\lambda_{i_p}\, \langle \d_{i_1}\ldots\d_{i_p}\rangle_c 
\nn\\
&=& -\frac{1}{2}\lambda_i\lambda_j\, \langle \d_i\d_j\rangle_c\, 
+\frac{(-i)^3}{3!}\,\lambda_i\lambda_j\lambda_k\, \langle \d_i\d_j\d_k\rangle_c \nn\\
&&+\frac{(-i)^4}{4!} \, \lambda_i\lambda_j\lambda_k\lambda_l\,
\langle \d_i\d_j\d_k\d_l\rangle_c
+\ldots\, ,
\ees
where $\d_i=\d(S_i)$ and the sum over $i,j,\ldots$ is understood.
This gives
\bees\label{WnNG}
&&W(\d_0;\d_1,\ldots ,\d_n;S_n)=\Dl
\\
&&\exp\left\{ i\sum_{i=1}^n\lambda_i\d_i+\sum_{p=2}^{\infty} \frac{(-i)^p}{p!}\,
 \sum_{i_1=1}^n\ldots \sum_{i_p=1}^n
\lambda_{i_1}\ldots\lambda_{i_p}\,
\langle \d_{i_1}\ldots\d_{i_p}\rangle_c
\right\}\, ,\nn
\ees
where
\be
\Dl \equiv
\inT\frac{d\lambda_1}{2\pi}\ldots\frac{d\lambda_n}{2\pi}\, ,
\ee
so
\bees\label{Pi0explicit}
&&\Pi_{\rm PS}(\d_0;\d_n;S_n)=\inT d\d_1\ldots d\d_{n-1}\,
\Dl\\
&&\exp\left\{ i\sum_{i=1}^n\lambda_i\d_i+\sum_{p=2}^{\infty} \frac{(-i)^p}{p!}\, 
\sum_{i_1=1}^n\ldots \sum_{i_p=1}^n
\lambda_{i_1}\ldots\lambda_{i_p}\,
\langle \d_{i_1}\ldots\d_{i_p}\rangle_c
\right\}\, ,\nn
\ees
and
\bees\label{Piexplicit}
&&\Pi_{\eps}(\d_0;\d_n;S_n)=\int_{-\infty}^{\d_c} d\d_1\ldots d\d_{n-1}\,
\Dl\\
&&\exp\left\{ i\sum_{i_1=1}^n\lambda_i\d_i+\sum_{p=2}^{\infty} \frac{(-i)^p}{p!}\, 
\sum_{i=1}^n\ldots \sum_{i_p=1}^n
\lambda_{i_1}\ldots\lambda_{i_p}\,
\langle \d_{i_1}\ldots\d_{i_p}\rangle_c
\right\}\, .\nn
\ees

\subsection{Perturbation over the markovian case}

As it was found in the classical paper 
by \cite{Bond},
when the density $\d(R)$  is smoothed with a sharp filter in momentum
space it satisfies the equation
\be\label{Langevin1}
\frac{\pa\d(S)}{\pa S} = \eta(S)\, ,
\ee
where here $S=\s^2(R)$ is the variance of the linear density field
smoothed on the scale $R$ and computed with a sharp filter in momentum
space, while
$\eta (S)$ satisfies 
\be\label{Langevin2}
\langle \eta(S_1)\eta(S_2)\rangle =\d (S_1-S_2)\, .
\ee
\Eqs{Langevin1} {Langevin2} are  formally the same as a Langevin equation with a Dirac-delta  noise $\eta(S)$.
In this case, as discussed in paper~I,
\be\label{corrkfilter}
\langle\d(S_i)\d(S_j)\rangle_c=
{\rm min}(S_i,S_j)\, ,
\ee
and for gaussian fluctuations, where all $n$-point connected
correlators with $n\geq 3$ vanish,
the probability density $W$ can be computed explicitly,
\be\label{W}
W^{\rm gm}(\d_0;\d_1,\ldots ,\d_n;S_n)=\frac{1}{(2\pi\eps)^{n/2}}\, 
\exp\left\{-\frac{1}{2\eps}\,
\sum_{i=0}^{n-1}  (\d_{i+1}-\d_i)^2\right\},
\ee
where the superscript ``gm'' (gaussian-markovian) reminds that this value of $W$ is computed for gaussian fluctuations, whose dynamics with respect to the smoothing scale is markovian.
Using this result, in paper~I we have shown that, in the continuum
limit, the distribution function $\Pi_{\eps=0}(\d;S)$, computed
with a sharp filter in
momentum space, satisfies a Fokker-Planck equation with the boundary
condition $\Pi_{\eps=0}(\d_c,S)=0$, and we have therefore recovered, from our
path integral approach, 
the standard result of excursion set
theory, 
\be\label{Pix0}
\Pi^{\rm gm}_{\eps=0} (\d_0;\d;S)
= \frac{1}{\sqrt{2\pi S}}\,
\[e^{-(\d-\d_0)^2/(2S)}-
e^{-(2\d_c-\d_0-\d)^2/(2S)}\]\, .
\ee
For a tophat filter in coordinate space, we have found 
in paper~I that
\eq{corrkfilter} is replaced by
\be\label{corrxfilter}
\langle\d(S_i)\d(S_j)\rangle_c=
{\rm min}(S_i,S_j) +\D(S_i,S_j)\, ,
\ee
where $S$ is now the  variance of the linear density field computed with 
 tophat filter in coordinate space.
We found that  (for the 
$\Lambda$CDM model used in paper~I) $\D(S_i,S_j)$ is very well approximated
by the simple analytic expression
\be\label{CTTaT}\label{approxDelta}
\D(S_i,S_j)\simeq \kappa\, \frac{S_i(S_j-S_i)}{S_j}\, ,
\ee
where $S_i\leq S_j$ (the value for $S_i>S_j$ is obtained by symmetry,
since $\D(S_i,S_j)=\D(S_j,S_i)$), and
$\kappa (R)$ is given in \eq{adiRlim}.
The term
${\rm min}(S_i,S_j)$ in \eq{corrxfilter} would be obtained if the dynamics where governed by the Langevin equation  \eq{Langevin1}, written with respect to the variance $S$ computed with the tophat filter in coordinate space, and
with a Dirac delta noise, and therefore describes the markovian part
of the dynamics. The term $\D(S_i,S_j)\equiv\D_{ij}$ is a correction
that reflects the fact that,  when one uses a tophat filter in
coordinate space, the underlying dynamics is
non-markovian. Observe that the full two-point correlator
(\ref{corrxfilter}) cannot be obtained from an underlying Langevin equation and, as a consequence, the probability distribution $\Pi_{\eps}(\d_0;\d_n;S_n)$ does not satisfy any local generalization of the Fokker-Planck equation, see the discussion below eq.~(83) of paper~I.  However, the formalism developed in paper~I allowed us to compute 
$\Pi_{\eps}(\d_0;\d_n;S_n)$  directly from its path integral representation,
\bees
&&\Pi_{\eps}(\d_0;\d_n;S_n) =
\int_{-\infty}^{\d_c} 
d\d_1\ldots d\d_{n-1}\,\Dl\nn\\
&&\times\exp\left\{
i\lambda_i\d_i -\frac{1}{2}
[{\rm min}(S_i,S_j) + \D(S_i,S_j)]
\lambda_i\lambda_j\right\}\, ,\label{Delta1}
\ees
by expanding perturbatively in $\D(S_i,S_j)$. The
zeroth-order term simply gives \eq{Pix0},
i.e. the standard excursion set result, with the variance of the
filter that we are using. The first correction is given by
\bees
&&\Pi^{{\D}1}_{\eps}(\d_0;\d_n;S_n) \equiv
\int_{-\infty}^{\d_c} 
d\d_1\ldots d\d_{n-1}\,\frac{1}{2}\sum_{i,j=1}^n \D_{ij}\pa_i\pa_j\nn\\
&&\times
\Dl\, 
\exp\left\{
i\sum_{i=1}^n\lambda_i\d_i -\frac{1}{2}\,\sum_{i,j=1}^n {\rm min}(S_i,S_j)
\lambda_i\lambda_j\right\}\label{Delta3}\\
&&=\frac{1}{2}\sum_{i,j=1}^n \D_{ij}\int_{-\infty}^{\d_c} 
d\d_1\ldots d\d_{n-1}\,\pa_i\pa_j
W^{\rm gm}(\d_0;\d_1,\ldots ,\d_n;S_n)
\, ,\nn
\ees
where we used  the notation
$\pa_i=\pa/\pa\d_i$ anf the identity
$\lambda e^{i\lambda x}=-i\pa_xe^{i\lambda x}$.
This quantity has been computed explicitly in Section~5.3 of paper~I,
and the corresponding 
result for the halo mass function is given by \eq{ourfI}. 
In this paper we will
perform a similar  computation for the correction induced by the
three-point function.

\section{Extension of  excursion  set theory to
non-Gaussian fluctuations}\label{sect:excNG}

If in \eq{Piexplicit}
we only retain the three-point correlator, and we use the tophat filter in
coordinate space, we have
\bees
&&\Pi_{\eps}(\d_0;\d_n;S_n) =
\int_{-\infty}^{\d_c} 
d\d_1\ldots d\d_{n-1}\,\Dl\label{Pi3}\\
&&\times
\exp\left\{i\lambda_i\d_i -\frac{1}{2}
[{\rm min}(S_i,S_j) + \D_{ij}]
\lambda_i\lambda_j
+\frac{(-i)^3}{6}\langle\d_i\d_j\d_k\rangle\lambda_i\lambda_j\lambda_k
\right\}. \nn
\ees
Expanding to first order, $\D_{ij}$ and 
$\langle\d_i\d_j\d_k\rangle$ do not mix, so we must compute
\be\label{Pi3full}
\Pi^{(3)}_{\eps}(\d_0;\d_n;S_n)\equiv 
-\frac{1}{6}\sum_{i,j,k=1}^n \langle\d_i\d_j\d_k\rangle
\int_{-\infty}^{\d_c} 
d\d_1\ldots d\d_{n-1}
\, \pa_i\pa_j\pa_k
W^{\rm gm}\, ,
\ee
where the superscript $(3)$ in $\Pi^{(3)}_{\eps}$
refers to the fact that this is the
contribution linear in the three-point correlator. 
In principle the expression given
in \eq{Pi3full}
can be computed using the formalism that we
developed in paper~I. 
In the continuum limit the triple sum over $i,j,k$ in \eq{Pi3full}
becomes a triple integral over  intermediate 
time variables $dS_i, dS_j, dS_k$,
each one integrated from zero to $S_n$, so the full result
is given by a triple time integral involving
$\langle\d(S_i)\d(S_j)\d(S_k)\rangle$, which is
not very illuminating.

Fortunately, such a full computation is not necessary either. 
Remember in fact that the non-Gaussianities are
particularly interesting at large masses. Large masses correspond to
small values of the  variance $S=\s^2(M)$.
Each of the integrals over $dS_i, dS_j, dS_k$
must therefore be performed over an interval $[0,S_n]$ that shrinks to
zero as $S_n\ra 0$. In this limit it is not necessary to take into 
account the exact functional form of
$\langle\d(S_i)\d(S_j)\d(S_k)\rangle$. Rather, to lowest order we
can replace it simply by $\langle\d^3(S_n)\rangle$. More generally,
we can expand
the three-point correlator in a triple
Taylor series around the point $S_i=S_j=S_k=S_n$.
We introduce the notation
\be
G_3^{(p,q,r)}(S_n)\equiv 
\[\frac{d^p}{dS_i^p}\frac{d^q}{dS_j^q}\frac{d^r}{dS_k^r}
\langle\d(S_i)\d(S_j)\d(S_k)\rangle\]_{S_i=S_j=S_k=S_n}\, .
\ee
Then
\bees\label{pqr}
&&\langle\d(S_i)\d(S_j)\d(S_k)\rangle
=\\
&&
\sum_{p,q,r=0}^{\infty} \frac{(-1)^{p+q+r}}{p!q!r!}
(S_n-S_i)^{p}(S_n-S_j)^{q}
(S_n-S_k)^{r}G_3^{(p,q,r)}(S_n)\, .\nn
\ees
We expect 
(and we will verify explicitly in the following) that terms
with more and more derivatives give contributions to the function
$f(\s)$, defined in \eq{dndMdef}, that are subleading in the limit of
small $\s$, i.e. for $\s/\d_c\ll 1$.
So, we expect that the leading contribution to the halo mass function 
will be given by the term in \eq{pqr} with $p=q=r=0$. At
next-to-leading order we must also include
the contribution of the terms in \eq{pqr}
with $p+q+r=1$, i.e. the three terms
$(p=1,q=0,r=0)$, $(p=0,q=1,r=0)$ and $(p=0,q=0,r=1)$, at
next-to-next-to-leading order
we must include the
contribution of the terms in \eq{pqr}
with $p+q+r=2$, and so on.

Observe  that, in a general theory, the functions
$G_3^{(p,q,r)}(S_n)$ with different values of $(p,q,r)$ are all
independent of each other; for instance,
\be
G_3^{(1,0,0)}(S_n) = \[\frac{d}{dS_i}
\langle\d(S_i)\d^2(S_n)\rangle\]_{S_i=S_n}\, ,
\ee
is in general not the same as 
\be
\frac{1}{3}\[ \frac{d}{dS}\langle\d^3(S)\rangle\]_{S=S_n}\, ,
\ee
so $G_3^{(1,0,0)}(S_n)$ cannot be written as a  derivative of 
$G_3^{(0,0,0)}(S_n)$. The terms $G_3^{(p,q,r)}(S_n)$ in
\eq{pqr} must all be treated as independent functions, that
characterize the most general non-gaussian theory (except, of course,
for the fact that $G_3^{(p,q,r)}(S_n)$ is symmetric under exchanges of
$p,q,r$).  
However, for the
purpose of organizing the expansion in leading term, subleading
terms, etc., we can
reasonably expect that, for small $S_n$
\be\label{ordering}
G_3^{(p,q,r)}(S_n)\sim
S_n^{-(p+q+r)}\langle\d^{3}(S_n)\rangle\, ,
\ee
i.e. each derivative $\pa/\pa S_i$,  when evaluated in $S_i=S_n$, gives a factor of order $1/S_n$.
This ordering  will be assumed when we present our
final result for the halo mass function below. However, our formalism
allows us  to compute each contribution 
separately, so our results below can be easily generalized in order to
cope with a different hierarchy between the various $G_3^{(p,q,r)}(S_n)$.

\subsection{Leading term}\label{sect:lead}

The leading term in $\Pi^{(3)}$ is
\be\label{Pi3appr}
\Pi^{(3,{\rm L})}_{\eps}(\d_0;\d_n;S_n)= 
-\frac{\langle\d_n^3\rangle}{6}\, 
\sum_{i,j,k=1}^n
\int_{-\infty}^{\d_c} 
d\d_1\ldots d\d_{n-1}\pa_i\pa_j\pa_k
W^{\rm gm}\, ,\nn
\ee
where the superscript ``${\rm L}$'' stands for ``leading''.
This expression
can  be computed very easily  by making use of a trick that we already 
introduced in  paper~I. Namely, we consider the derivative 
of $\Pi^{\rm gm}_{\eps}$ with
respect to $\d_c$ (which, when we use the notation
$\Pi^{\rm gm}_{\eps}(\d_0;\d_n;S_n)$,
is not written explicitly in the list of
variable on which  $\Pi^{\rm gm}_{\eps}$ depends, but of course
enters as upper integration limit in
\eq{defPi}). The first derivative with respect to $\d_c$
can be written as (see eq.~(B8) of paper~I)
\be
\frac{\pa}{\pa \d_c}\Pi^{\rm gm}_{\eps} (\d_0;\d_n;S_n)=
\sum_{i=1}^{n-1}
\int_{-\infty}^{\d_c} d\d_1\ldots d\d_{n-1}\, 
\pa_iW^{\rm gm}\, ,
\ee
since, when  $\pa/\pa \d_c$ acts on the upper integration
limit of the integral over $d\d_i$, it produces 
$W(\d_1,\ldots,\d_i=\d_c,\ldots, \d_n;S_n)$, which is the same as the integral of
$\pa_iW$ with respect to $d\d_i$ from $\d_i=-\infty$ to $\d_i=\d_c$. Similarly 
\be\label{pa2xc}
\frac{\pa^2}{\pa \d_c^2}\Pi^{\rm gm}_{\eps} (\d_0;\d_n;S_n)=
\sum_{i,j=1}^{n-1}
\int_{-\infty}^{\d_c} d\d_1\ldots d\d_{n-1}\, 
\pa_i\pa_jW^{\rm gm}\, ,
\ee
see eqs.~(B9) and (B10) of paper~I. In the same way we find that
\be\label{pa3xc}
\frac{\pa^3}{\pa \d_c^3}\Pi^{\rm gm}_{\eps} (\d_0;\d_n;S_n)=
\sum_{i,j,k=1}^{n-1}
\int_{-\infty}^{\d_c} d\d_1\ldots d\d_{n-1}\, 
\pa_i\pa_j\pa_k W^{\rm gm}\, .
\ee
The right-hand side of this identity is not yet equal to the quantity
that appears in \eq{Pi3appr}, since there the sums run up to $n$ while
in \eq{pa3xc} they only run up to $n-1$. However, what we need is not
really $\Pi^{(3)}_{\eps}(\d_0;\d_n;S_n)$, but rather its integral over
$d\d_n$, which is the quantity that enters in \eq{calFTPi}. 
Then we consider
\bees\label{Pi3appr2}
&&\int_{-\infty}^{\d_c}d\d_n\, \Pi^{(3,{\rm L})}_{\eps}(\d_0;\d_n;S_n)=
-\frac{1}{6}\, \langle\d_n^3\rangle\nn\\
&&\times\sum_{i,j,k=1}^n
\int_{-\infty}^{\d_c} 
d\d_1\ldots d\d_{n-1}d\d_{n}\pa_i\pa_j\pa_k
W^{\rm gm}\, ,
\ees
and we can now
use the identity
\bees
&&\sum_{i,j,k=1}^{n}
\int_{-\infty}^{\d_c} d\d_1\ldots d\d_{n-1}d\d_n\,
\pa_i\pa_j\pa_k W^{\rm gm}\nn\\
&=&
\frac{\pa^3}{\pa \d_c^3}
\int_{-\infty}^{\d_c} d\d_1\ldots d\d_{n-1}d\d_n\,
W^{\rm gm}\nn\\
&=& \frac{\pa^3}{\pa \d_c^3}\int_{-\infty}^{\d_c}d\d_n\,\
\Pi^{\rm gm}_{\eps} (\d_0;\d_n;S_n)\, ,\label{ident3der}
\ees
so
\be\label{Pi3appr3}
\int_{-\infty}^{\d_c}d\d_n\, \Pi^{(3,{\rm L})}_{\eps}(\d_0;\d_n;S_n)=
-\frac{\langle\d_n^3\rangle}{6}
\frac{\pa^3}{\pa \d_c^3}\int_{-\infty}^{\d_c}d\d_n\,
\Pi^{\rm gm}_{\eps} (\d_0;\d_n;S_n)\, .
\ee
From \eq{Pix0}, setting for simplicity $\d_0=0$, 
\be\label{Pigaueps0}
\Pi^{\rm gm}_{\eps=0} (\d_0=0;\d_n;S_n)
= \frac{1}{\sqrt{2\pi S_n}}\,
\[e^{-\d_n^2/(2S_n)}-
e^{-(2\d_c-\d_n)^2/(2S_n)}\]\, .
\ee
Inserting this into \eq{Pi3appr3} we immediately find the result in
the continuum limit,
\be\label{Pi3finite}
\int_{-\infty}^{\d_c}d\d_n\, \Pi^{(3,{\rm L})}_{\eps=0}(0;\d_n;S_n)
=\frac{\langle\d_n^3\rangle}{3\sqrt{2\pi}\, S_n^{3/2}}
\( 1-\frac{\d_c^2}{S_n}\)
\, e^{-\d_c^2/(2S_n)}\, .
\ee
We now insert this result into \eqs{calFTPi}{deffs}
and we express the result in terms of the normalized skewness 
\be\label{defskew}
{\cal S}_3(\s)\equiv \frac{1}{S^2}\langle\d^3(S)\rangle\, .
\ee
Putting the contribution of $\Pi^{(3,{\rm L})}$ 
together with the gaussian contribution,
we find
\bees\label{fsR1}
&&f(\s)= \(\frac{2}{\pi}\)^{1/2}\, 
\frac{\d_c}{\s}\, \, e^{-\d_c^2/(2\s^2)}\\
&&\times \left\{ 1+\frac{\s^2}{6\d_c}
\[ {\cal S}_3(\s)
\(\frac{\d_c^4}{\s^4}-\frac{2\d_c^2}{\s^2}-1\)
+\frac{d{\cal S}_3}{d\ln\s}\, \(\frac{\d_c^2}{\s^2}-1\)\]\right\}\, .\nn
\ees
Remarkably, this agrees exactly with the result obtained by
\cite{LoVerde}, performing an Edgeworth expansion of the
non-Gaussian generalization of Press-Schechter  theory,
see \eq{RNG:LoVerde}.

However, the fact that a naive non-Gaussian generalization
of PS theory gives the
same result that we have obtained from the  non-Gaussian
generalization of excursion set theory
(at least to leading order for small $\s/\d_c$; we will
see below that the subleading term gets corrections) is somewhat
accidental, as can be realized as follows.
In the
sum over $i,j,k$ of $\pa_i\pa_j\pa_k$ in \eq{Pi3appr}, it is useful
to separate
the contribution  with $i=j=k=n$ from the rest. Recall  that in PS theory
the upper integration limit for the variables
$d\d_1, \ldots ,d\d_{n-1}$ 
is $+\infty$ rather than $\d_c$
(which reflects the fact that in PS theory one looks at the
probability that, at a given smoothing radius, the smoothed density is
above threshold, regardless of whether it was already above threshold
for some larger smoothing radius). If in \eq{Pi3appr} we replaced the
upper integration limit $\d_c$ with $+\infty$, a derivative $\pa_i$ with
$i<n$ would integrate by parts to zero. 
The terms
where at least one of the indices $i,j$ or $k$ is strictly smaller
than $n$ therefore  have no counterpart in PS
theory.
The term where all indices $i,j,\ldots$ are equal to $n$, in contrast,
are local terms, which depends only on the cumulants rather than on
the correlators at different points, and 
that can have a correspondence with PS theory. In the
gaussian case, it is just such a local term that gives back the PS
result, together with the factor of two that in PS theory was added by
hand. Formally, this comes from the fact that, in the gaussian case,
the excursion set
probability distribution is the difference between the original PS gaussian
and an ``image'' gaussian, and these two terms give contributions that
add up when computing the first crossing rate.

In the case of the three-point correlator the situation is however different.
Denoting by
$\Pi^{(3,{\rm La})}$ the  contribution to
$\Pi^{(3,{\rm L})}$ obtained by setting $i=j=k=n$ in 
\eq{Pi3appr}, we have
\be
\Pi^{(3,{\rm La})}_{\eps=0}(0;\d_n;S_n) =-\frac{1}{6}
\langle\d_n^3\rangle\pa_n^3\Pi^{\rm gm}_{\eps=0}(0;\d_n;S_n)
\, ,
\ee
and therefore 
\be\label{Ia}
\int_{-\infty}^{\d_c}d\d_n\, 
\Pi^{(3,{\rm La})}_{\eps=0}(0;\d_n;S_n)
=-\frac{1}{6}
\langle\d_n^3\rangle
\[\pa_n^2\Pi^{\rm gm}_{\eps=0}(0;\d_n;S_n)\]_{\d_n=\d_c}=0\, .
\ee
This result is in a sense
surprising. 
Since PS theory gives a wrong normalization factor, missing a factor of
two in the gaussian case, and a factor close to two  in the
non-Gaussian case, what is done in the literature when one uses PS theory
is to take the PS result and multiply it  by hand by a factor of two
(or, for non-Gaussian fluctuations, close to two), 
assuming that this would come out from a proper
treatment of the cloud-in-cloud problem, i.e. from excursion set
theory. We see however that this is not at all the case. In excursion
set theory $\Pi^{\rm gm}_{\eps=0}$ is a difference of two gaussians,
see \eq{Pix0}, so all its derivative with respect to $\d_n$ of odd
order, evaluated in $\d_n=\d_c$ are twice as large as for a single
gaussian, but the function itself, as well as 
all its derivative with respect to $\d_n$ of even
order, evaluated in $\d_n=\d_c$, are zero, i.e. the contribution from
the second gaussian cancels the first contribution, rather than
adding up. Since in \eq{Ia} appears the second derivative
of $\Pi^{\rm gm}_{\eps=0}$ in $\d_n=\d_c$, this term vanishes. We
therefore see that the logic behind the use of PS theory for non-Gaussian
fluctuations, namely (1): compute with a naive extension
of PS theory to non-Gaussian fluctuations and  (2): multiply the
result by hand by
a ``fudge factor'' $\simeq 2$, assuming that it would come out from 
a solution of the cloud-in-cloud problem, is not justified. 
For the contribution linear in the three-point correlator
$\langle\d_n^3\rangle$, this ``fudge factor'' is actually zero, and
the result comes entirely from terms with at least one derivative
$\pa_i$ with $i<n$, which have no counterpart in a non-Gaussian 
extension of PS theory. 
Above we have computed the excursion set theory result performing at
once the sum over $i,j,k$, using the trick given in \eq{ident3der}.
In appendix~\ref{app:B} we  compute separately the 
terms in the sum over $i,j,k$ with one or more indices equal to $n$, 
and we check that they give back
\eq{Pi3finite}.

In  Sections~\ref{sect:appro2} and \ref{sect:appro3}
we will
compute the  corrections to \eq{fsR1}
to next-to-leading and to
next-to-next-to-leading order.
We also need to take into account that the barrier must be treated as
diffusing, see paper~II, and we must include
the corrections due to the tophat filter
in coordinate space. This will be done in 
Section~\ref{sect:diffandfilter}.

Before leaving this section we observe that, in the approximation
in which the correlators are replaced by the cumulants, 
the effects of the higher-order correlators can also be
computed very simply. For instance, the effect of the four-point
function $\langle\d_n^4\rangle$ is obtained using
\bees
&&\sum_{i,j,k,l=1}^{n}
\int_{-\infty}^{\d_c} d\d_1\ldots d\d_{n-1}d\d_n\,
\pa_i\pa_j\pa_k\pa_l W^{\rm gm}\nn\\
&=& \frac{\pa^4}{\pa \d_c^4}\int_{-\infty}^{\d_c}d\d_n\,\
\Pi^{\rm gm}_{\eps} (\d_0;\d_n;S)\, .
\ees

\subsection{The next-to-leading term}
\label{sect:appro2}

Using \eqs{Pi3full}{pqr}, at  next-to-leading order we get
\bees\label{Pi3apprcNL}
&&\int_{-\infty}^{\d_c}  d\d_n\, \Pi^{(3,{\rm NL})}_{\eps}(\d_0;\d_n;S_n)
=\frac{1}{2}\,G_3^{(1,0,0)}(S_n)\\
&&\times \sum_{i=1}^n (S_n-S_i)\sum_{j,k=1}^n
\int_{-\infty}^{\d_c} 
d\d_1\ldots d\d_{n-1}d\d_n\pa_i\pa_j\pa_k
W^{\rm gm}\, ,\nn
\ees
where the superscript ``NL'' in $\Pi^{(3,{\rm NL})}_{\eps}$
stands for next-to-leading,
and we used the fact that the three terms
$(p=1,q=0,r=0)$, $(p=0,q=1,r=0)$ and $(p=0,q=0,r=1)$ give the same
contribution. 
We now use the same trick as before to eliminate
$\sum_{j,k=1}^n\pa_j\pa_k$ in favor of $\pa^2/\pa \d_c^2$, 
\bees\label{Pi3apprc}
&&\int_{-\infty}^{\d_c}  d\d_n\Pi^{(3,{\rm NL})}_{\eps}(\d_0;\d_n;S_n)=
\frac{1}{2}\,
G_3^{(1,0,0)}(S_n)\nn\\
&&\times\sum_{i=1}^n (S_n-S_i)\frac{\pa^2}{\pa \d_c^2}
\int_{-\infty}^{\d_c} 
d\d_1\ldots d\d_{n-1}d\d_n\pa_i
W^{\rm gm}\, .\nn
\ees
The remaining path integral can be computed using the technique developed in
paper~I, namely we write
\bees
&&\int_{-\infty}^{\d_c} 
d\d_1\ldots d\d_{n-1}d\d_n\, \pa_iW^{\rm gm}\\
&=&\int_{-\infty}^{\d_c} 
d\d_1\ldots d\d_{n-1}d\d_n\,
W(\d_0;\d_1,\ldots, \d_i=\d_c,\ldots ,\d_n;S_n)
\, ,\nn
\ees
and we use
\bees\label{facto}
&&W^{\rm gm}(\d_0;\d_1,\ldots, \d_{i-1}, \d_c, \d_{i+1}, \ldots ,\d_n; S_n)\\
&&= 
W^{\rm gm}(\d_0;\d_1,\ldots, \d_{i-1}, \d_c;S_i)
W^{\rm gm}(\d_c; \d_{i+1}, \ldots ,\d_n; S_n-S_i)\, ,\nn
\ees
so
\bees\label{WPiPi}
&&\int_{-\infty}^{\d_c}d\d_1\ldots d\d_{i-1}
\int_{-\infty}^{\d_c}  d\d_{i+1}\ldots  d\d_{n-1}d\d_n\nn\\
&&\times W^{\rm gm}(\d_0;\d_1,\ldots, \d_{i-1}, \d_c;S_i)
W^{\rm gm}(\d_c; \d_{i+1}, \ldots ,\d_n; S_n-S_i)\nn\\
&&=\Pi^{\rm gm}_{\eps}(\d_0;\d_c;S_i)
\int_{-\infty}^{\d_c}  d\d_n\,  \Pi^{\rm gm}_{\eps}(\d_c;\d_n;S_n-S_i)\, .
\ees
Recalling from  paper~I that
\be\label{Pigammafinal}
\Pi^{\rm gm}_{\eps} (\d_0;\d_c;S)=
\sqrt{\eps}\, \frac{1}{\sqrt{\pi}}\, 
\frac{\d_c-\d_0}{S^{3/2}} e^{-(\d_c-\d_0)^2/(2S)} +{\cal O}(\eps)
\ee
and
\be\label{Pigammafinalbis}
\Pi^{\rm gm}_{\eps} (\d_c;\d_n;S)=
\sqrt{\eps}\, \frac{1}{\sqrt{\pi}}\, 
\frac{\d_c-\d_n}{S^{3/2}} e^{-(\d_c-\d_n)^2/(2S)} +{\cal O}(\eps)\, ,
\ee
we see that the factors $\sqrt{\eps}$ in
$\Pi^{\rm gm}_{\eps} (\d_0;\d_c;S)$ and in $\Pi^{\rm gm}_{\eps}
(\d_c;\d_n;S)$ combine with $\sum_i$ to produce an integral over $dS_i$,
and 
\bees\label{Pi3apprcinter}
&&\int_{-\infty}^{\d_c}  d\d_n\, \Pi^{(3,{\rm NL})}_{\eps}(\d_0;\d_n;S_n)
=\frac{1}{2\pi}\, G_3^{(1,0,0)}(S_n)\nn\\
&&\times
\int_0^{S_n}dS_i\, \frac{1}{S_i^{3/2} (S_n-S_i)^{1/2}}\\
&&\times\frac{\pa^2}{\pa \d_c^2}\, 
\[ \d_c e^{-\d_c^2/(2S_i)}\int_{-\infty}^{\d_c}d\d_n\,
(\d_c-\d_n) \exp\left\{-\frac{(\d_c-\d_n)^2}{2(S_n-S_i)}\right\} \]
\, .\nn
\ees
The integral over $d\d_n$ is easily performed writing
\be
(\d_c-\d_n) \exp\left\{-\frac{(\d_c-\d_n)^2}{2(S_n-S_i)}\right\}
=(S_n-S_i)\pa_n\exp\left\{-\frac{(\d_c-\d_n)^2}{2(S_n-S_i)}\right\}
\, ,
\ee
so it just gives $(S_n-S_i)$.
Carrying out the second derivative
with respect to $\d_c$ and the remaining elementary integral over
$dS_i$  we  get
\be\label{Pi3NL}
\int_{-\infty}^{\d_c}  d\d_n\, \Pi^{(3,{\rm NL})}_{\eps}(\d_0;\d_n;S_n)
=\frac{1}{\sqrt{2\pi}}\, \frac{G_3^{(1,0,0)}(S_n)}{S_n^{1/2}}
\, e^{-\d_c^2/(2S_n)}\, .
\ee
We now define
\be
{\cal U}_3(\s)\equiv \frac{3 G_3^{(1,0,0)}(S)}{S}\, ,
\ee
where as usual $S=\s^2$.
When the ordering given in \eq{ordering} holds, ${\cal U}_3(\s)$ is of the same order
as the normalized skewness ${\cal S}_3(\s)$ given in \eq{defskew}.
Computing the contribution to $f(\s)$ from
\eq{Pi3NL} and putting it together with \eq{fsR1} we finally find
\be\label{fsR1NL}
f(\s)= \(\frac{2}{\pi}\)^{1/2}\, 
\frac{\d_c}{\s}\, \, e^{-\d_c^2/(2\s^2)}
\[ 1+\frac{\s^2}{6\d_c} h_{\rm NG}(\s) \]\, ,
\ee
where
\bees
h_{\rm NG}(\s)&=&
\frac{\d_c^4}{\s^4} {\cal S}_3(\s)
-\frac{\d_c^2}{\s^2} 
\( 2  {\cal S}_3(\s)+\, {\cal U}_3(\s) -\frac{d{\cal S}_3}{d\ln\s}
 \)\nn\\
&&-\( {\cal S}_3(\s) +\, {\cal U}_3(\s) +\frac{d{\cal S}_3}{d\ln\s}
 +\frac{d{\cal U}_3}{d\ln\s} \)\label{defhsigma}
\, .
\ees
We have ordered the terms in $h_{\rm NG}(\s)$ according to their importance in 
the limit of small $\s/\d_c$ assuming, according to \eq{ordering},
that ${\cal U}_3(\s)$ is of the same order as ${\cal S}_3(\s)$.
The  leading order is given by
$(\d_c/\s)^4{\cal S}_3(\s)$
and, as we have seen, it comes only from $\Pi^{(3,{\rm L})}$. The
next-to-leading order in $h_{\rm NG}(\s)$
is given by the terms  proportional to
$(\d_c/\s)^2$, and we see that it is affected by 
the terms with $p+q+r=1$ in
the expansion of \eq{pqr}.
The terms in $h_{\rm NG}(\s)$ which are ${\cal O}(1)$ with respect to 
the large parameter $\d_c/\s$
are next-to-next-to-leading order corrections and, if we wish to
include them,  we must for consistency include also the contribution from 
the terms with $p+q+r=2$ in
the expansion of
\eq{pqr}. We compute them in the next subsection.

Observe also that 
typically
${\cal S}_3$ depends very weakly on the smoothing
scale $R$ and hence on $\s$.  For
instance, in $\fnl$-theories 
it changes only by a factor $\simeq 3$ as $R$ is changed
by a factor 100, from $0.1\, {\rm Mpc}/h$ to $10\, {\rm Mpc}/h$, 
see \cite{MVJ}. Therefore, even if parametrically
$d{\cal S}_3/d\ln\s$ has the same power-law behavior as 
${\cal S}_3$, its prefactor will typically be numerically small.

\subsection{The next-to-next-to-leading term}
\label{sect:appro3}

Using \eqs{Pi3full}{pqr} and keeping the terms with $p+q+r=2$ we find two
kind of contributions. The first has $(p=2,q=r=0)$, with a combinatorial
factor of three and the second has $(p=q=1, r=0)$, again
with a combinatorial
factor of three. We denote the
contribution to $\Pi^{(3)}$ at next-to-next-to-leading (NNL) order by 
$\Pi^{(3,{\rm NNL})}$, and the two separate contribution
with $(p=2,q=r=0)$ and with $(p=q=1, r=0)$ as
$\Pi^{(3,{\rm NNLa})}$ and $\Pi^{(3,{\rm NNLb})}$, respectively.
Thus,
\bees\label{Pi3NNLa}
&&\int_{-\infty}^{\d_c}  d\d_n\, \Pi^{(3,{\rm NNLa})}_{\eps}(\d_0;\d_n;S_n)
=-\frac{1}{4}\,G_3^{(2,0,0)}(S_n)\\
&&\times \sum_{i=1}^n (S_n-S_i)^2\sum_{j,k=1}^n
\int_{-\infty}^{\d_c} 
d\d_1\ldots d\d_{n-1}d\d_n\pa_i\pa_j\pa_k
W^{\rm gm}\, ,\nn
\ees
and
\bees\label{Pi3NNLb}
&&\int_{-\infty}^{\d_c}  d\d_n\, \Pi^{(3,{\rm NNLb})}_{\eps}(\d_0;\d_n;S_n)
=-\frac{1}{2}\,G_3^{(1,1,0)}(S_n)\\
&&\times \sum_{i,j=1}^n (S_n-S_i)(S_n-S_j)\sum_{k=1}^n
\int_{-\infty}^{\d_c} 
d\d_1\ldots d\d_{n-1}d\d_n\pa_i\pa_j\pa_k
W^{\rm gm}\, .\nn
\ees
The first term is straightforward to compute. We use again the trick
of eliminating $\sum_{j,k=1}^n\pa_j\pa_k$ in favor of $\pa^2/\pa
\d_c^2$, and we proceed just as in Section~\ref{sect:appro2}. The
result is
\bees\label{Pi3NNLa2}
&&\int_{-\infty}^{\d_c}  d\d_n\, \Pi^{(3,{\rm NNLa})}_{\eps}(\d_0;\d_n;S_n)
=-\frac{3}{4\pi }\,G_3^{(2,0,0)}(S_n)\nn\\
&&\times
\[ \sqrt{2\pi}\,  S_n^{1/2} e^{-\d_c^2/(2S_n)}
-\pi \d_c\, {\rm Erfc}\(\frac{\d_c}{\sqrt{2S_n}}\)\]\, ,
\ees
where Erfc is the complementary error function.

The computation of \eq{Pi3NNLb} is more complicated, but can be
performed with the formalism that we have developed in paper~I, see
in particular appendix~B of paper~I. The factor $\sum_{k=1}^n\pa_k$ is
eliminated as usual in favor of $\pa/\pa \d_c$.
We also observe that, in \eq{Pi3NNLb}, the terms with $i=n$ or $j=n$
do not contribute, because of the factor
$(S_n-S_i)(S_n-S_j)$, and we separate the sum over $i,j$ into the term
with $i=j$ and twice the term with $i<j$. The first term is
\bees\label{Pi3NNLb2}
I_1&\equiv& -\frac{1}{2}\,G_3^{(1,1,0)}(S_n)
\sum_{i=1}^{n-1} (S_n-S_i)^2\nn\\
&&\times\frac{\pa}{\pa \d_c}
\int_{-\infty}^{\d_c} 
d\d_1\ldots d\d_{n-1}d\d_n\pa^2_i
W^{\rm gm}\, ,
\ees
and the second is
\bees
I_2&\equiv&-G_3^{(1,1,0)}(S_n)
\sum_{i=1}^{n-2}\sum_{j=i+1}^{n-1} 
(S_n-S_i)(S_n-S_j)\nn\\
&&\times\frac{\pa}{\pa \d_c}
\int_{-\infty}^{\d_c} 
d\d_1\ldots d\d_{n-1}d\d_n\pa_i\pa_j
W^{\rm gm}\, .\label{Pi3NNLb3}
\ees
As we discussed in detail in paper~I, quantities such as
the right-hand side of \eq{Pi3NNLb} are finite in the continuum limit
$\eps\ra 0$, as it is obvious physically, and as we checked explicitly
in solvable examples in paper~I. However, when we split the sum
over the indices $i,j$ into two separate parts, such as those given in
\eqs{Pi3NNLb2}{Pi3NNLb3}, these are separately divergent
in the continuum limit, and the divergence cancels when we sum them
up. It is therefore necessary to regularize them carefully, and
separate them into a divergent part 
and the finite part. Since we know that the divergent terms
must cancel,
we can simply extract from each term the finite part,
disregarding the divergences. This is the finite part prescription
discussed and tested in detail in paper~I. We will denote by
${\cal FP}$ this procedure of extracting the finite part from terms
such as (\ref{Pi3NNLb2}) and (\ref{Pi3NNLb3}).

The computation of the finite part of (\ref{Pi3NNLb2}) is
basically identical to the one that we already performed  in
appendix~B of paper~I, see in particular eqs.~(B13)--(B15) and
(B29) there, and
the result is that this term diverges as $1/\sqrt{\eps}$ with no
finite part, so
\be
{\cal FP}\sum_{i=1}^{n-1} (S_n-S_i)^2\frac{\pa}{\pa \d_c}
\int_{-\infty}^{\d_c} 
d\d_1\ldots d\d_{n-1}d\d_n\pa^2_i
W^{\rm gm} =0\, .
\ee
The computation (\ref{Pi3NNLb3}) is also completely analogous to the
computation of the ``memory-of-memory'' term performed in 
appendix~B of paper~I, see in particular eqs.~(B17)--(B28) and
(B30) there, except that we now have a factor
$(S_n-S_i)(S_n-S_j)$ in the integrals over $dS_i$ and $dS_j$.
We can then repeat basically the same steps as detailed
in appendix~B of paper~I, and we find
\bees
{\cal FP}[I_2]&=&-\frac{G_3^{(1,1,0)}(S_n)}{\pi\sqrt{2\pi}}
{\cal FP}
\int_0^{S_n}dS_i\,
\frac{(S_n-S_i)}{S_i^{3/2}}\, \(1-\frac{\d_c^2}{S_i}\) e^{-\d_c^2/2S_i}
\nn\\
&&\times\int_{S_i}^{S_n}dS_j\, 
\frac{(S_n-S_j)^{1/2}}{(S_j-S_i)^{3/2}}
\,\exp\left\{-\frac{a^2}{2(S_j-S_i)}\right\}\, ,
\ees
where $a=\sqrt{\a\eps}$ and $\a$ is a numerical constant which appears
when the sum over $j$ is replaced by an integral over $dS_j$, see
eqs.~(B20)-(B24) of paper~I.
The integral over 
$dS_j$ can be computed  writing $t_n=S_n-S_i$,
$t_j=S_j-S_i$, and using  the identity
\bees
&&\int_{0}^{t_n}dt_j\, \frac{(t_n-t_j)^{1/2}}{t_j^{1/2}}
\exp\left\{-\frac{a^2}{2t_j}\right\}\nn\\
&=&\frac{\sqrt{2\pi}}{2a}\, 
\[ 2 t_n^{1/2} e^{-a^2/(2 t_n)} -a\sqrt{2\pi}\, \,
{\rm Erfc}\(\frac{a}{\sqrt{2 t_n}}\)\]\, ,\label{Pi3NNLb4}
\ees
which is proved in the same way as  eqs.~(115) and (116) of paper~I.
In this equation $a=\sqrt{\a\eps}$ goes to zero in the continuum
limit.  In the limit
$a\ra 0$ the above result displays a term divergent as $1/a$, i.e. as
$1/\sqrt{\eps}$, which must cancel the divergence coming from
(\ref{Pi3NNLb2}), plus a term which is finite as $a\ra 0$, which can
be extracted from \eq{Pi3NNLb4} recalling that
${\rm Erfc}(0)=1$, so
\be
{\cal FP}\int_{S_i}^{S_n}dS_j\, 
\frac{(S_n-S_j)^{1/2}}{(S_j-S_i)^{3/2}}
\,\exp\left\{-\frac{a^2}{2(S_j-S_i)}\right\}=-\pi\, .
\ee
Computing the remaining integral over $dS_i$, which is finite and
elementary, we find
\bees
&&\int_{-\infty}^{\d_c}  d\d_n\, \Pi^{(3,{\rm NNLb})}_{\eps}(\d_0;\d_n;S_n)
=-\frac{2}{\pi }\, G_3^{(1,1,0)}(S_n)\nn\\
&&\times
\[ \sqrt{2\pi}\,  S_n^{1/2} e^{-\d_c^2/(2S_n)}
-\pi \d_c\, {\rm Erfc}\(\frac{\d_c}{\sqrt{2S_n}}\)\]\, .
\ees
Putting together this result and \eq{Pi3NNLa2} we end up with
\bees\label{Pi3NNLbfinal}
&&\int_{-\infty}^{\d_c}  d\d_n\, \Pi^{(3,{\rm NNL})}_{\eps}(\d_0;\d_n;S_n)\nn\\
&=&-\frac{1}{2\pi }\,
\(\frac{3}{2}G_3^{(2,0,0)}(S_n)+4 G_3^{(1,1,0)}(S_n)\)\nn\\
&&\times
\[ \sqrt{2\pi}\,  S_n^{1/2} e^{-\d_c^2/(2S_n)}
-\pi \d_c\, {\rm Erfc}\(\frac{\d_c}{\sqrt{2S_n}}\)\]\, .
\ees
We now introduce
the function
\be
{\cal V}_3(\s)\equiv  \frac{9}{2} G_3^{(2,0,0)}(S)+12 G_3^{(1,1,0)}(S)
\, .
\ee
According to \eq{ordering}, ${\cal V}_3(\s)$ is parametrically 
of the same order as ${\cal S}_3(\s)$ and ${\cal U}_3(\s)$, as 
$\s\ra 0$. We can now compute the contribution of this term to the
function $h_{\rm NG}(\s)$ using \eqss{calFTPi}{deffs}{fsR1NL}. 
Retaining only the terms that
contribute up to ${\cal O}(1)$ in $\d_c/\s$,  
we find that the function $h_{\rm NG}(\s)$ is modified to
\bees
h_{\rm NG}(\s)&=&
\frac{\d_c^4}{\s^4} {\cal S}_3(\s)
-\frac{\d_c^2}{\s^2} 
\( 2  {\cal S}_3(\s)+\, {\cal U}_3(\s) -\frac{d{\cal S}_3}{d\ln\s}
 \)\nn\\
&&-\( {\cal S}_3(\s)+\, {\cal U}_3(\s)+{\cal V}_3(\s)
 +\frac{d{\cal S}_3}{d\ln\s}
  +\frac{d{\cal U}_3}{d\ln\s}\)\nn\\
&&+{\cal O}\(\frac{\s^2}{\d_c^2}\)
\, .\label{defhsigmaV3}
\ees
This is the complete result for the halo mass function, up to 
NNL order in the small parameter $\s^2/\d_c^2$.

\subsection{The effects of the diffusing barrier and of the 
filter}\label{sect:diffandfilter}

Until now we have worked with a barrier with a fixed height $\d_c$
and we neglected the corrections due to the filter.
We now include the modifications due to the fact that the height of
the barrier diffuses stochastically, as discussed in paper~II, and
also the corrections due to the filter.

To compute the non-Gaussian term proportional to 
the three-point correlator with the diffusing barrier we recall, from
paper~II, that the first-passage time problem of a particle obeying a
diffusion equation with diffusion coefficient $D=1$, in the presence of a
barrier that moves stochastically with diffusion coefficient $D_B$,
can be mapped into the first-passage time problem of a particle with
effective diffusion coefficient $(1+D_B)$, and fixed barrier. This can be
reabsorbed into a rescaling of the ``time'' variable $S\ra (1+D_B)S=S/a$,
and therefore $\s\ra \s/\sqrt{a}$. At the same time 
the three-point correlator must be rescaled according to
$\langle\d_n^3\rangle\ra a^{-3/2}\langle\d_n^3\rangle$ since,
dimensionally, $\langle\d_n^3\rangle$ is the same as $S^{3/2}$
(if we perform dimensional analysis as discussed below eq.~(A10) of paper~I), which means that
${\cal S}_3\ra a^{1/2}{\cal S}_3$, and similarly for
the functions ${\cal U}_3$ and ${\cal V}_3$.\footnote{In principle 
we should
also shift the argument $\s$ of ${\cal S}_3$,
${\cal U}_3$ and ${\cal V}_3$. However, 
${\cal S}_3$ depends very weakly on the smoothing
scale $R$ and hence on $\s$.  For
instance, in $\fnl$-theories 
it changes only by a factor $\simeq 3$ as $R$ is changed
by a factor 100, from $0.1\, {\rm Mpc}/h$ to $10\, {\rm Mpc}/h$, 
see \cite{MVJ}. In
most situation, we can then neglect the rescaling of 
the argument of ${\cal S}_3$, and we expect that the same holds for
${\cal U}_3$ and ${\cal V}_3$.}
Then \eqs{fsR1NL}{defhsigmaV3} become
\be\label{fsR1NLdiff}
f(\s)= \(\frac{2}{\pi}\)^{1/2}\, 
\frac{a^{1/2}\d_c}{\s}\, \, e^{-a\d_c^2/(2\s^2)}
\[ 1+\frac{\s^2}{6a^{1/2}\d_c} h_{\rm NG}(\s) \]\, ,
\ee
where
\bees
h_{\rm NG}(\s)&=&
\frac{a^2\d_c^4}{\s^4} {\cal S}_3(\s)
-\frac{a\d_c^2}{\s^2} 
\( 2  {\cal S}_3(\s)+\, {\cal U}_3(\s) -\frac{d{\cal S}_3}{d\ln\s}
 \)\nn\\
&&-\( {\cal S}_3(\s) +\, {\cal U}_3(\s)+{\cal V}_3(\s)
+\frac{d{\cal S}_3}{d\ln\s}
  +\frac{d{\cal U}_3}{d\ln\s}
 \)\nn\\
&&+{\cal O}\(\frac{\s^2}{\d_c^2}\)
\, .
\label{defhsigmaV3diff}
\ees
We see that the terms depending on
the skewness ${\cal S}_3(\s)$ and its derivative 
coincide with those given in \eq{RNG:LoVerde}, if we identify
$\d_{\rm eff}$ with $a^{1/2}\d_c$. Observe, from \eq{a079},
that our prediction $a^{1/2}\simeq  0.89$ is  in remarkable
agreement with the value $a^{1/2}\simeq 0.86$ proposed by
\cite{Grossi} from the fit to the $N$-body simulations (see however 
footnote~\ref{foot:a}).

We have therefore derived, from a first principle computation,
\eq{RNG:LoVerde}, which was proposed in  
\cite{LoVerde} and in \cite{Grossi} using a mixture of
heuristic theoretical arguments (the use of a non-Gaussian extension
of PS theory, rather than of the
excursion set theory) and a calibration of parameters from the fit to
the data
of the $N$-body simulations 
(the replacement  $\d_c\ra 0.86\d_c$), and we have improved it
including the effect of the
functions ${\cal U}_3(\s)$ and ${\cal V}_3(\s)$, which
are absent in \cite{LoVerde} and cannot be obtained from any
naive extension of PS theory,
which from the beginning contains only the cumulants, rather 
than the full correlation functions at different smoothing radii.

The  term in \eq{defhsigmaV3diff} which is dominant for small $\s$ is
the same as that of  both \eqs{RNG:MVJ}{RNG:LoVerde}, and  
appears to fit well the data of the
$N$-body simulations~\citep{Grossi}. Given the size of  the  error bars  of the
non-Gaussian $N$-body simulations
(see e.g. Fig.~6 and 7 of \cite{Grossi}), it is probably 
difficult for the moment  to test the
subleading terms in \eq{defhsigmaV3diff}, and in particular 
to see the effect
of the functions  ${\cal U}_3(\s)$ and ${\cal V}_3(\s)$.

As a final ingredient, we must add the effect of the tophat filter
function in coordinate space.
When the non-gaussianities are not
present, these are given by \eq{ourf}.
More generally, even the
non-Gaussian corrections must be computed using the propagator
$[{\rm min}(S_i,S_j) + \D_{ij}]$ in \eq{Pi3}, so we will apply the
same correction factor found for the gaussian part 
also to the non-Gaussian term, and we end up with
\bees\label{fsR5} 
f(\s)&=& (1-\tilde{\kappa})\, \(\frac{2}{\pi}\)^{1/2}\, 
\frac{a^{1/2}\d_c}{\s}\, \, e^{-a\d_c^2/(2\s^2)}
\[ 1+\frac{\s^2}{6a^{1/2}\d_c} h_{\rm NG}(\s) \]\nn\\
&&+\frac{\tilde{\kappa}}{\sqrt{2\pi}}\,
\frac{a^{1/2}\d_c}{\s}\,
 \G\(0,\frac{a \d_c^2}{2\s^2}\)\, ,
\ees
with $h_{\rm NG}(\s)$ still given by \eq{defhsigmaV3diff}.
More generally, also the term proportional to the incomplete Gamma function
could get non-Gaussian corrections, which in principle can be computed
evaluating perturbatively a ``mixed'' term proportional to 
\be
\D_{ij}\langle\d_k\d_l\d_m\rangle\pa_i\pa_j\pa_k\pa_l\pa_m
\ee
in \eq{Pi3}. However we saw in paper~I that 
in the large mass limit, where the
non-Gaussianities are important, the term proportional to the incomplete
Gamma function is subleading, so we will neglect the non-Gaussian
corrections to this subleading term.\footnote{Furthermore, one must be
aware of the fact that the term
proportional to $h_{\rm NG}(\s)$ might in general receive
corrections from the tophat filter that have not exactly the same form
as that of the gaussian term. Again, in principle these can be obtained
by computing
the term  proportional to 
$\D_{ij}\langle\d_k\d_l\d_m\rangle\pa_i\pa_j\pa_k\pa_l\pa_m$ in the
expansion of the path integral.}

The relative weight of the correction due to the filter proportional
to the incomplete Gamma function, and of the non-Gaussian corrections depends
on the value of $\s$ and, of course, on the value of the three-point
correlator, i.e. of ${\cal S}_3$. In $\fnl$ theory
${\cal S}_3$ increases very weakly with the mass, i.e. as $\s\ra 0$.
In the low-$\s$ (i.e. large mass) limit we can use the asymptotic
expansion of the incomplete Gamma function for large $z$,  
$\Gamma(0,z)\simeq z^{-1} e^{-z}$, and we see that, asymptotically, the 
term in the second line of \eq{fsR5} depends on $\s$  as
$\s \exp\{-a\d_c^2/(2\s^2)\}$, and therefore is small compared to both
the leading and next-to-leading term in the non-Gaussian corrections, 
which overall behaves as 
$\s^{-3}{\cal S}_3(\s)\exp\{-a\d_c^2/(2\s^2)\}$
and 
$\s^{-1}{\cal S}_3(\s)\exp\{-a\d_c^2/(2\s^2)\}$, respectively. When
this asymptotic behavior sets in depends, of course, on the numerical
value of ${\cal S}_3$ so, in $\fnl$-theory, on the value of the 
$\fnl$ parameter. In any case, given a measure of $f(\s)$, either from
galaxy surveys or from $N$-body simulations with non-Gaussian initial
conditions, the prediction (\ref{fsR5}) allows us to disentangle the
effects due to the filter from the physically interesting 
effects due to primordial non-Gaussianities.

\section{Conclusions}\label{sect:concl}

To conclude this series of three papers, we summarize 
the main results that we obtained and, at the price of some
repetition, we collect here the most important
formulas that are scattered in the text. Our aim was to compute the
halo mass function, i.e. the number density $n(M)dM$ of dark matter halos 
with mass between $M$ and $M+dM$, both for gaussian and non-Gaussian
primordial density fluctuations. This can be written as 
\be
\frac{dn(M)}{dM} = f(\s) \frac{\bar{\rho}}{M^2} 
\frac{d\ln\s^{-1} (M)}{d\ln M}\, ,
\ee
and the issue is to compute the function $f(\s)$. Our final result can
be written as
\bees\label{concl1}
f(\s)&=& (1-\tilde{\kappa})\, \(\frac{2}{\pi}\)^{1/2}\, 
\frac{a^{1/2}\d_c}{\s}\, \, e^{-a\d_c^2/(2\s^2)}
\[ 1+\frac{\s^2}{6a^{1/2}\d_c} h_{\rm NG}(\s) \]\nn\\
&&+
\frac{\tilde{\kappa}}{\sqrt{2\pi}}\,
\frac{a^{1/2}\d_c}{\s}\,
 \G\(0,\frac{a \d_c^2}{2\s^2}\)\, ,
\ees
where $\Gamma(0,z)$ is the incomplete Gamma function.
Three distinct physical effects are taken into account in this result. 

One is the fact that
we have treated the threshold for gravitational collapse
as a stochastic variable that fluctuates around an average
value, which is $\d_c\simeq 1.686$ for the spherical collapse model,
and is a rising function of $\s$ for the ellipsoidal collapse
model. As discussed in paper~II, this is a way of taking into account, at least at an effective level, part of the complexity of a realistic process of halo formation, which is missed in the simple spherical or ellipsoidal collapse model. Furthermore, the stochasticity of the barrier reflects  uncertainties in the operative definition of what is a dark matter halo. The inclusion of a diffusing barrier
gives rise to the constant  $a$ in the above result. This constant enters also in the exponential, thereby
modifying dramatically the  behavior predicted by PS theory. 
Our prediction is $a\simeq 0.80$, i.e. $\sqrt{a}\simeq 0.89$,
which gives a remarkable agreement
with the data from $N$-body simulations. For instance
\cite{Grossi}, from the fit to the $N$-body simulation, find 
$\sqrt{a}\simeq 0.86$.

A second effect included in \eq{concl1}
is that 
we have properly  accounted for the fact that the comparison with
the data, whether observational or from $N$-body simulations,
requires the use of a tophat filter
function in coordinate space. In the classical paper of \cite{Bond},
using a tophat filter in {\em momentum} space, the computation of
$f(\s)$ was reduced to a first-passage time problem for a quantity
that obeys a Langevin equation, and therefore the underlying dynamics
is markovian. When one considers a different filter function, the
dynamics becomes non-markovian and therefore the problem is much more
complicated. Basically, this is the issue that 
for a long time blocked further
analytical progress on this problem. In paper~I of this series we have
developed a formalism in which the problem is formulated in terms of a
path integral with boundaries, and non-markovian corrections can be
computed perturbatively. In \eq{concl1} this effect enters through
the constant $\tilde{\kappa}$, defined as $\tilde{\kappa}=a\kappa$
with $\kappa$ given by \eq {adiRlim}.

The third effect, which was the subject of the present paper, is the
inclusion of the non-Gaussianities. These are contained in the
function $h_{\rm NG}(\s)$. Using the path integral technique developed
in paper~I, we have computed it to leading, next-to-leading and
next-to-next-to-leading order in the  parameter $\s^2/\d_c^2$, which
is small for large halo masses,  where one can hope to
see the effect of non-Gaussianities on the halo mass function. Our result is
\bees
h_{\rm NG}(\s)&=&
\frac{a^2\d_c^4}{\s^4} {\cal S}_3(\s)
-\frac{a\d_c^2}{\s^2} 
\( 2  {\cal S}_3(\s)+\, {\cal U}_3(\s) -\frac{d{\cal S}_3}{d\ln\s}
 \)\nn\\
&&-\( {\cal S}_3(\s) +\, {\cal U}_3(\s)+{\cal V}_3(\s)
+\frac{d{\cal S}_3}{d\ln\s}
  +\frac{d{\cal U}_3}{d\ln\s}
 \)\nn\\
&&+{\cal O}\(\frac{\s^2}{\d_c^2}\)
\, .\label{concl3}
\ees
The functions ${\cal S}_3(\s)$, ${\cal U}_3(\s)$ and ${\cal V}_3(\s)$
are defined in terms of the three-point correlator of the smoothed
density field $\langle \d(S_1) \d(S_2) \d(S_3)\rangle$ and of its
derivatives, as follows,
\bees
{\cal S}_3&=&\frac{1}{S^2}\, \langle \d^3(S)\rangle\, ,\\
{\cal U}_3&=&\frac{3}{S}\,
\[\frac{d}{dS_1}\langle \d(S_1)\d^2(S)\rangle\]_{S_1=S}\, ,\\
{\cal V}_3&=&\frac{9}{2}
\[\frac{d^2}{dS_1}\langle \d(S_1)\d^2(S)\rangle\]_{S_1=S}\nn\\
&&+12\[\frac{d}{dS_1}\frac{d}{dS_2}
\langle \d(S_1)\d(S_2)\d(S)\rangle\]_{S_1=S_2=S}\, ,
\ees
and we prefer to write them as functions of $\s=\sqrt{S}$.

Our result has passed to a good accuracy various comparisons with
numerical results. First of all, one can 
study numerically what happens in the excursion set theory, with fixed (rather
than diffusing) barrier and tophat filter in coordinate space, by 
performing a
Monte Carlo realization of the
first-crossing distribution of excursion set theory,
obtained by integrating numerically a
Langevin equation with a colored noise. This was recently performed in
detail in
\cite{RKTZ} (see also \cite{Bond}). In this limit our analytical
result is obtained from
\eq{concl1} setting $a=1$ (since the barrier is taken as fixed in the
Monte Carlo simulation) and $h_{\rm NG}(\s)=0$, i.e. we are testing the
effect of $\kappa$. Comparing our results in paper~I with 
Fig.~4 of \cite{RKTZ} we find very good agreement.
This is a first useful test of our technique.

Using \eq{concl1} with  $a\simeq 0.80$ (obtained by reading the diffusion coefficient
of the barrier $D_B$ from $N$-body simulations, and using our prediction $a=1/(1+D_B)$)
and with
$\kappa$ given in \eq {adiRlim}, and setting 
$h_{\rm NG}(\s)=0$, we can compare our result with the mass function found in $N$-body
simulations with gaussian initial conditions. The comparison is shown in
Figs.~6 and 7 of paper~II.
For all values of $\s^{-1}\geq 0.3$ the discrepancy between our analytic result and the Tinker et al. fit to  the same
$N$-body simulation is 
smaller than  $20\%$, and for $\s^{-1}\geq 1$ it is smaller than
$10\%$. Considering that our result comes from an analytic model of halo formation with no tunable parameter (the parameter $a$ is fixed once $D_B$ is given, and we do not have the right to tune it),  while the Tinker et al. fitting formula
is  simply a fit to the data with four free parameters, we think that this result is quite encouraging. The numerical accuracy
is actually the best that one could have hoped for, considering for instance that we have neglected  second-order non-markovian corrections.

Finally, our prediction for the function $h_{\rm NG}(\s)$ can be
tested against $N$-body
simulations with non-Gaussian initial conditions. To leading order in
the small $\s$ limit, our result reduces to that proposed by 
\cite{LoVerde} and \cite{MVJ} using  non-Gaussian extensions of PS
theory, and it has been found in \cite{Grossi} that this formula
reproduces very well the data, see in particular their 
Figs.~6 and~7. 
The size of the error bars is probably still too large for
discriminating between different forms of the subleading term.

\vspace{5mm}

\noindent We thank Sabino Matarrese 
for useful discussions.
The work
of MM is supported by the Fond National Suisse. 
The work of AR is supported by 
the European Community's Research Training Networks 
under contract MRTN-CT-2006-035505.

\vspace{2cm}
\appendix

\section{A. Extension of
Press-Schechter theory to non-Gaussian fluctuations}\label{sect:PS}

As we repeatedly emphasized, the really interesting quantity for
comparison with experimental data from galaxy surveys, and with
$N$-body simulations, is the distribution function $\Pi_{\eps}$, that
generalizes excursion set theory to non-Gaussian fluctuations. The
function $\Pi_{\rm PS}$ defined in \eq{defPi0}, where the integrations
over the variables $d\d_i$
run up to $+\infty$ rather than up to $\d_c$, not only suffers from the
fact that it predicts that only a fraction of the total mass of the
Universe finally ends up in virialized objects (the infamous factor of
two that Press-Schechter where forced to introduce by hand) but also
misses all the subtle correlations between different scales which
are just one of the characteristic features of
non-Gaussianities. For this reason, in the body of this paper we
concentrated on the computation of $\Pi_{\eps}$. Still, it is
interesting to see how our path integral formalism reproduces PS
theory and generalizes it to non-Gaussian theories. 
We discuss the issue in this appendix. 
In particular, we
will see that, even in the non-Gaussian case, $\Pi_{\rm PS}$ 
satisfies a differential equation which is local in ``time'', the
Kramers-Moyal equation, and which generalizes the Fokker-Planck
equation. It is interesting to contrast this result with what
happens for $\Pi_{\eps}$ which instead, as discussed in paper~I, does not
satisfy any local diffusion-like equation.

With our ``microscopic'' formalism based on the path integral,
it is very easy to derive   PS theory and to extend it
to non-Gaussian fluctuations. Simply, in \eq{Pi0explicit} each
integral over $d\d_i$, with $1\leq i\leq n-1$, produces a factor
$2\pi\d_D(\lambda_i)$, which allows us to perform trivially
all the integrals over $d\lambda_i$ with $i<n$.
Denoting the residual variable $\lambda_n$ by $\lambda$ and setting for
notational simplicity $\d_0=0$ (the general result is recovered with
$\d\ra \d-\d_0$),
\eq {Pi0explicit} becomes
\be\label{PSnongauresult3}
\Pi_{\rm PS}(\d_0=0;\d;S)=\inT\frac{d\lambda}{2\pi}\, 
\exp\{ i\lambda \d +
\sum_{p=2}^{\infty} \frac{(-i\lambda)^p}{p!} \mu_p(S) 
\}\, .
\ee
When all $\mu_p$ with $p\geq 3$ vanish, the integral gives a gaussian
and we get back the standard PS result,
\be\label{PiPSmu2}
\Pi^{\rm PS} (\d_0=0;\d;S)=
\frac{1}{(2\pi S)^{1/2}}\, 
e^{-\d^2/2S}
\, ,
\ee
since, by definition $\mu_2(S)=S$, where $S$ is   the variance  computed with the filter
function of our choice.
\Eq{PSnongauresult3} generalizes  PS theory to arbitrary
non-Gaussian theories.\footnote{A word of caution is 
necessary when one considers \eq{PSnongauresult3}
with correlators $\mu_p$ with $p\geq 4$. For instance, keeping only 
$\mu_2$, $\mu_3$ and $\mu_4$, one is faced with an integral
that diverges, since $\mu_4(S)=\langle\d^4(S)\rangle> 0$. 
The correct statement is
that $\Pi_{\rm PS}(\d_0;\d;S)$ is given, order by order, by
the expansion of \eq{PSnongauresult3} in powers of $\mu_4$.
However, the
expansion in powers of $\mu_4$ is only an  asymptotic series, which
can be used to approximate the true result up to a finite  order in 
$\mu_4$, but diverges if we keep an infinite number of terms.
If instead the highest cumulant that we include 
in \eq{PSnongauresult3} is 
$\mu_6$, the integral converges because $(-i)^6\mu_6 =-\mu_6 <0$, while
the integral diverges  again if the highest cumulant that we include 
in \eq{PSnongauresult3} is 
$\mu_8$,  since $(-i)^8\mu_8 =+\mu_8 >0$, and so on. 
Anyhow,
the whole issue of the full resummation of the contributions
of the $\mu_4$ or higher-order
correlators is  physically irrelevant. These
correlators are in general computed using phenomenological
parametrization of the non-Gaussianities, such as $\fnl$-theory, that
are meant to be a useful description
of the true non-Gaussianities
only to leading, and at most next-to-leading 
order in $\fnl$, so in general only the first few 
terms in the series  makes sense physically.}
Observe that \eqs{PSnongauresult3}{PiPSmu2} hold independently of
the filter function used, and the $\mu_p$ are the cumulants computed
with the filter function in which one is interested.

\Eq{PSnongauresult3} is a well-known result in the theory of stochastic
processes (see e.g. ~\cite{Risken}), and it was applied to $\fnl$-theory in
\cite{MVJ}. 
Using this expression, the usual strategy in the literature
is to compute ${\cal F}(S)$ using
\be\label{F(M)PS}
{\cal F}_{\rm PS}(S)=\frac{\pa}{\pa T}
\int_{\d_c}^{\infty}dx\, \Pi_{\rm PS}(\d_0;\d; S)\, ,
\ee
and to multiply by hand by a fudge factor
$\simeq 2$ to ensure the proper normalization. As we have shown
in the discussion below \eq{Ia},  this
multiplication by a fudge factor
is not justified for non-Gaussianities. Still, let us
discuss from the mathematical point of view the properties
of the function $\Pi_{\rm PS}$, in order to contrast them with
the excursion set theory distribution function $\Pi_{\eps}$.

First of all, it is instructive to rederive the expression
(\ref{PSnongauresult3}) for
$\Pi_{\rm PS}$, with generic filter and generic
non-Gaussian theory in an alternative way, using the technique
developed in paper~I for computing
the effect of the correction $\D_{ij}$ to the two-point function,
see \eqs{Delta1}{Delta3}.
To compute 
$\Pi_{\rm PS}(\d_0;\d;S)$ when the two-point correlator 
$\langle \d_i\d_j\rangle_c$ is generic, rather than equal to 
${\rm min}(S_i,S_j)$, and in the presence of the higher-order correlators,
we write
\be
\langle \d_i\d_j\rangle_c = {\rm min}(S_i,S_j) +
[\langle \d_i\d_j\rangle_c -  {\rm min}(S_i,S_j) ]
\equiv \eps A_{ij} +\eps B_{ij}\, .\label{sepAB}
\ee
Observe that  $\eps B_{nn}=\mu_2(S)-S=0$.
We then expand the exponential 
in \eq{Pi0explicit}
in powers of
$\eps B_{ij}$ and of the higher-order correlators, 
\bees
\Pi_{\rm PS}(\d_0;\d_n;S_n)&=&
\int_{-\infty}^{\infty}
d\d_1\ldots d\d_{n-1}\,\Dl
\[ 1 -\frac{1}{2} \,\sum_{i,j=1}^n \lambda_i\lambda_j \eps B_{ij}
+\frac{(-i)^3}{3!}\,\sum_{i,j,k=1}^n
\lambda_i\lambda_j\lambda_k\, \langle \d_i\d_j\d_k\rangle_c 
+\ldots\] \,e^{
i\sum_{i=1}^n\lambda_i\d_i 
-\frac{1}{2} \, \sum_{i,j=1}^n\lambda_i\lambda_j \eps A_{ij}} \nn\\
&=&
\int_{-\infty}^{\infty} d\d_1\ldots d\d_{n-1}\,\label{ExtePi0}
\[ 1 +\frac{1}{2} \, \sum_{i,j=1}^n\eps B_{ij}\pa_i\pa_j 
-\frac{1}{3!}\,
\sum_{i,j,k=1}^n \langle \d_i\d_j\d_k\rangle_c \, \pa_i\pa_j\pa_k
+\ldots\]\Dl\,
e^{
i\sum_{i=1}^n\lambda_i\d_i 
-\frac{1}{2} \, \sum_{i,j=1}^n\lambda_i\lambda_j \eps A_{ij}}\, ,
\ees
where $\pa_i=\pa/\pa \d_i$. 
The derivatives $\pa_i$ contribute only
when the index $i=n$, otherwise we have a total derivative with
respect to an integration variable, and the corresponding 
boundary terms at $\d=\pm\infty$
term vanish. Here it is crucial that one integrates up to
$+\infty$. When we rather consider $\Pi_{\eps}$, instead of
$\Pi_{\rm PS}$, the upper integration limit is $\d_c$ and we remain
with complicated and non-local boundary terms, compare e.g. with
eq.~(83) of paper~I.
For $\Pi_{\rm PS}$ however this boundary term is absent and
\bees
\Pi_{\rm PS}(\d_0;\d_n;S_n)&=&
\[ 1 -\frac{1}{3!}\, \langle \d_n^3\rangle_c \, \pa_n^3
+\ldots\]\int_{-\infty}^{\infty}
d\d_1\ldots d\d_{n-1}\,
W^{\rm gm}(\d_0;\d_1,\ldots, \d_i,\ldots ,\d_n;S_n)\nn\\
&=&
\[ 1-\frac{1}{3!}\, \langle \d_n^3\rangle_c \, \pa_n^3
+\ldots\]
\Pi^{0,\rm gau}(\d_0;\d;S)
\, .
\ees
Since the derivative $\pa_n=\pa/\pa \d_n$ does not act on the
correlators $\langle\d^p_n\rangle$ (which are functions of
$S_n$, but not of $\d_n$), the expansion in the square brackets
can be exponentiated back, and we can write
\be\label{ngPS1}
\Pi_{\rm PS}(\d_0;\d;S)=
e^{\hat{K}_{\rm NG}}\Pi^{0,\rm gau}(\d_0;\d;S)\, ,
\ee
where (using now $\d_0$ generic)
\be
\Pi^{0,\rm gau}(\d_0;\d;S)=\frac{1}{(2\pi S)^{1/2}}\,
e^{-(\d-\d_0)^2/(2S)}\, ,
\ee
and
the differential operator $\hat{K}_{\rm NG}$ is given by
\be
\hat{K}_{\rm NG}=
\sum_{p=3}^{\infty}\, \frac{(-1)^p}{p!}\mu_p(S)\frac{\pa^p}{\pa\d^p} 
\label{ngPS2}\label{defhatK3}
\, .
\ee
To prove
the equivalence of \eqs{ngPS1}{PSnongauresult3} 
we write \eq{PSnongauresult3} as
\be\label{Pi0WNG}
\Pi_{\rm PS}(\d_0=0;\d;S)=\inT\frac{d\lambda}{2\pi}\, 
\exp\left\{ i\lambda \d -\frac{1}{2}\mu_2(S)\lambda^2 + 
W_{\rm NG}(\lambda)\right\}\, ,
\ee
with
\be
W_{\rm NG}(\lambda)=
\sum_{p=3}^{\infty} \frac{(-i\lambda)^p}{p!} \mu_p(S) 
\, ,
\ee
and we expand the exponential in powers of $W_{\rm NG}(\lambda)$. Using
$\lambda^p e^{i\lambda x} = (-i\pa_x)^p e^{i\lambda x}$
we see that
$W_{\rm NG}(\lambda)e^{i\lambda x}=W(-i\pa_x) e^{i\lambda x}$,
and the same holds for any power of $W_{\rm NG}(\lambda)$, so 
\be
\exp\{W_{\rm NG}(\lambda)\} e^{i\lambda x} =
\exp\{W_{\rm NG}(-i\pa_x)\} e^{i\lambda x}
=e^{\hat{K}_{\rm NG}}e^{i\lambda x}\, .
\ee
Therefore \eq{Pi0WNG} becomes
\be
\Pi_{\rm PS}(\d_0=0;\d;S)=e^{\hat{K}_{\rm NG}}
\inT\frac{d\lambda}{2\pi}\, 
\exp\left\{ i\lambda \d -\frac{1}{2}\mu_2(S)\lambda^2\right\}\, ,
\ee
which agrees with \eq{ngPS1}.
So, the distribution function $\Pi_{\rm PS}$ that gives the extension
of the PS formalism to non-Gaussian fluctuations can be written
equivalently in the integral form (\ref{PSnongauresult3}) or in the
differential form (\ref{ngPS1}), with ${\hat{K}_{\rm NG}}$
given by \eq{ngPS2}.

It is  interesting to observe that the function
$\Pi_{\rm PS}(\d_0=0;\d;S)$ obeys a local differential equation, both in the
gaussian and in the non-Gaussian case.
Consider first a gaussian theory with a generic
filter function, so $\mu_p(S)=0$ for $p\geq 3$. 
In order to see exactly where enters the
difference between integrating up to $\d_c$,  as in $\Pi_{\eps}$ and
integrating up to $+\infty$,
we start from the definition\
\be\label{3defPi0}
\Pi_{\rm PS}(\d_0;\d_n;S_n) =
\inT d\d_1\ldots d\d_{n-1}\,\Dl\,
e^{
i\lambda_i\d_i -\frac{1}{2}
\langle\d_i\d_j\rangle
\lambda_i\lambda_j}\, ,
\ee
where the sum over $i,j=1,\ldots ,n$ is understood,
and we derive a differential equation satisfied by $\Pi_{\rm PS}$,
by taking
the derivative  with respect to $S_n$,
\be\label{deriFP1}
\frac{\pa\Pi_{\rm PS}}{\pa S_n}=
\(-\frac{1}{2}\, 
\frac{\pa \langle\d_k\d_l\rangle_c}{\pa S_n}\)
\inT d\d_1\ldots d\d_{n-1}\,
\Dl\,
\lambda_k\lambda_l
\exp\left\{
i\lambda_i\d_i -\frac{1}{2}
\langle\d_i\d_j\rangle
\lambda_i\lambda_j\right\}\, .\nn
\ee
Again, using $\lambda\exp\{i\lambda x\}=-i\pa_x\exp\{i\lambda x\}$, 
inside the integral we can replace
$\lambda_k\ra  -i\pa_k$
and $\lambda_l\ra -i\pa_l$. Since  we integrate over
$d\d_1, \ldots d\d_{n-1}$, but not over $d\d_n$,
if $k\neq n$ the term
$\pa_k$, when integrated over $d\d_k$, is a total derivative and gives
zero, because at the boundaries  $\d_k=\pm\infty$ the integrand vanishes
exponentially, and the only contribution comes from $k=n$.
Similarly, also $\pa_l$ contributes only when $l=n$. 
This is the step that does not go through for $\Pi_{\eps}$, when the
integration is only up to $\d_c$, and a complicated boundary term
arises, see eq.~(83) of paper~I.
Therefore, since $\langle\d_n^2\rangle =S_n$,
we get a Fokker-Planck  equation
\be
\label{FPcoloredmu2}
\frac{\pa\Pi_{\rm PS}}{\pa S}=\frac{1}{2} 
\frac{\pa^2\Pi_{\rm PS}}{\pa \d^2}\, ,
\ee
whose solution, on the line $-\infty<\d<\infty$, is indeed given by
\eq{PiPSmu2}. 
\Eq{FPcoloredmu2} can be generalized to the
non-Gaussian case
using the integral form of the solution (\ref{PSnongauresult3})
and taking the time derivative,
\be\label{PSKM1}
\frac{\pa}{\pa S}\Pi_{\rm PS}=\sum_{p=2}^{\infty} 
\frac{\dot{\mu}_p(S)}{p!}
\inT\frac{d\lambda}{2\pi}\, (-i\lambda)^p
\exp\left\{ i\lambda \d +
\sum_{q=2}^{\infty} \frac{(-i\lambda)^q}{q!} \mu_q(S) 
\right\}\, ,
\ee
where $\dot{\mu}_p=d\mu_p/dS$.
Inside the integral we can replace $(i\lambda)^p e^{i\lambda x}$ by
$\pa_x^p e^{i\lambda x}$, so
\be\label{derKM}
\frac{\pa\Pi_{\rm PS}}{\pa S}
=\sum_{p=2}^{\infty}\frac{(-1)^p}{p!} \dot{\mu}_p(S) 
\frac{\pa^p\Pi_{\rm PS}}{\pa \d^p}\, .
\ee
This equation is called the Kramers-Moyal (KM) equation or ``the stochastic
equation'', and is well known in the theory of stochastic
processes (\cite{Stratonovich}, \cite{Risken}).

In conclusion we have seen that, independently of choice of filter
function, $\Pi^{\rm PS}$ satisfies a local
differential equation both in the gaussian and in the non-Gaussian
case. In the gaussian case it satisfies
the FP equation (\ref{FPcoloredmu2}), while in the non-Gaussian case
it satisfies the Kramer-Moyal equation (\ref{derKM}). As we already
saw in paper~I, this is not true for the distribution function
$\Pi_{\eps}$ of the excursion set formalism, unless one use a sharp
filter in momentum space and the theory is gaussian. Already for
gaussian theory and a different filter, we saw in eq.~(83) of paper~I
that the equation satisfied by $\Pi_{\eps}$, besides the Fokker-Planck
operator, contains
complicated non-local terms, coming from boundary terms at the
upper integration limit $\d_c$. The same happens, of course, when we
include the non-Gaussianities.

\section{B. Term-by-term computation of $\Pi^{(3,{\rm L})}$}\label{app:B}

In Section~\ref{sect:lead} we showed that
\be\label{Pi3finiteapp}
\int_{-\infty}^{\d_c}d\d_n\, \Pi^{(3,{\rm L})}_{\eps=0}(0;\d_n;S_n)
=\frac{1}{3}\, \frac{\langle\d_n^3\rangle}{\sqrt{2\pi}\, S_n^{3/2}}
\( 1-\frac{\d_c^2}{S_n}\)
\, e^{-\d_c^2/(2S_n)}\, .
\ee
Our derivation  used the fact that we could replace the
sum over $i,j,k$ of $\pa_i\pa_j\pa_k$ in \eq{Pi3appr} by $\pa^3/\pa
\d_c^3$. It is instructive to reproduce this result by evaluating
separately the various terms in the sum, and using the perturbative
formalism of paper~I. 
We then split 
$\sum_{i,j,k=1}^n$ into the following terms:
(a) $i=j=k =n$. (b) $i<n, j=k=n$. (c) $i=j <n, k=n$.
(d) $i<j<n, k=n$. (e) $\sum_{i,j,k=1}^{n-1}$, each one with its own
combinatorial factor. We denote the corresponding contributions to
$\Pi^{(3,{\rm L})}$ as $\Pi^{(3,{\rm La})}$, $\Pi^{(3,{\rm Lb})}$,
etc. and, for simplicity, we also use the notation
\be
I^{(a)}=\int_{-\infty}^{\d_c}d\d_n\, 
\Pi^{(3,{\rm La})}_{\eps=0}(0;\d_n;S_n)\, ,
\ee
and so on. As in the computation of the term
proportional to $\D_{ij}\pa_i\pa_j$ in paper~I, we  find that the
various contributions in this computation can be
separately divergent in the continuum limit 
$\eps\ra 0$, while their sum is  finite, as it clear
physically, and as we already know from our derivation in
Section~\ref{sect:lead}.  Indeed, 
the great virtue of the derivation performed in 
Section~\ref{sect:lead},  using the trick
of replacing the sum over $\pa_i\pa_j\pa_k$ with derivatives
with respect to $\d_c$
as in \eq{Pi3appr3}, is that it directly
gives the sum over all combination of indices, thus providing directly
the total finite result,  
and bypassing all problems of divergences that appear if
one compute separately the terms corresponding to different
combination of indices. 

The separate terms can however be computed using the technique developed in
paper~I, with the finite part prescription.
The term $I^{(a)}$ has already been computed in
\eq{Ia}, and we have seen that it vanishes.
The term $(b)$ is obtained setting $j=k=n$ in \eq{Pi3appr}, and taking
into account a combinatorial factor of three corresponding to the
three way of choosing which index, among $(i,j,k)$, is not equal to
$n$, so
\be\label{Pi3apprapp1}
\Pi^{(3,{\rm Lb})}_{\eps}(\d_0;\d_n;S_n)= 
-\frac{\langle\d_n^3\rangle}{2}\, 
\sum_{i=1}^{n-1}\pa_n^2
\int_{-\infty}^{\d_c} 
d\d_1\ldots d\d_{n-1}\pa_i
W^{\rm gm}
=-\frac{\langle\d_n^3\rangle}{2}\, 
\sum_{i=1}^{n-1}\pa_n^2 
\[\Pi^{\rm gm}_{\eps}(\d_0;\d_c;S_i)\Pi^{\rm gm}_{\eps}(\d_c;\d_n;S_n-S_i)\]
\, .
\ee
Using \eqs{Pigammafinal}{Pigammafinalbis},
\be\label{Ib}
I^{(b)}=-\frac{\langle\d_n^3\rangle}{2\pi}\, \[\pa_n
\int_0^{S_n}dS_i\, \frac{\d_c(\d_c-\d_n)}{S_i^{3/2}(S_n-S_i)^{3/2}}
\exp\left\{-\frac{\d_c^2}{2S_i}-\frac{(\d_c-\d_n)^2}{2(S_n-S_i)}
\right\}\]_{\d_n=\d_c}
=\frac{\langle\d_n^3\rangle}{\sqrt{2\pi}\, S_n^{3/2}}
\( 1-\frac{\d_c^2}{S_n}\)
\, e^{-\d_c^2/(2S_n)}\, .
\ee
The term $(c)$ gives
\be
\Pi^{(3,{\rm Lc})}_{\eps}(\d_0;\d_n;S_n) 
=-\frac{\langle\d_n^3\rangle}{2}
\sum_{i=1}^{n-1}\pa_n\int_{-\infty}^{\d_c} 
d\d_1\ldots d\d_{n-1}\pa^2_i
W^{\rm gm}\, ,
\ee
so
\be
I^{(c)}
=-\frac{\langle\d_n^3\rangle}{2}
\sum_{i=1}^{n-1}\[ \pa_i
\(\Pi^{\rm gm}_{\eps}(\d_0;\d_c;S_i)
\Pi^{\rm gm}_{\eps}(\d_c;\d_n;S_n-S_i)\)
\]_{\d_n=\d_c}
\ee
This expression is analogous to the one that has already been computed
in eqs.~(B13)-(B16) of paper~I, and it is purely divergent as
$1/\sqrt{\eps}$, with no finite part, so ${\cal FP}[I^{(c)}]=0$. 

The term $(d)$ is slightly more complicated, since it requires the
$\a$ regularization described in appendix~B of paper~I for extracting
the finite part. Setting $i<j<n, k=n$ in \eq{Pi3appr} and taking into
account a combinatorial factor of six, we get
\be\label{Pi3LcappB}
\Pi^{(3,{\rm Lc})}_{\eps}(\d_0;\d_n;S_n)
=-\langle\d_n^3\rangle \pa_n
\sum_{i=1}^{n-2}\sum_{j=i+1}^{n-1}  
\Pi^{\rm gm}_{\eps}(\d_0;\d_c;S_i)
\Pi^{\rm gm}_{\eps}(\d_c;\d_c;S_j-S_i)
\Pi^{\rm gm}_{\eps}(\d_c;\d_n;S_n-S_j)\, .
\ee
We use  \eqs{Pigammafinal}{Pigammafinalbis}, together with
$\Pi^{\rm gm}_{\eps}(\d_c;\d_c;S)=\eps/(\sqrt{2\pi}\, S^{3/2})$, see
eq.~(112) of paper~I, and we get
\be
I^{(d)}=-\frac{\langle\d_n^3\rangle}{\pi\sqrt{2\pi}}
\lim_{\d_n\ra \d_c^-}
\int_{0}^{S_n}dS_i\int_{S_i}^{S_n}dS_j\,
\frac{\d_c(\d_n-\d_c)}{S_i^{3/2} (S_j-S_i)^{3/2}(S_n-S_j)^{3/2}}
\,\exp\left\{-\frac{\d_c^2}{2S_i}-\frac{A^2}{2(S_j-S_i)}
-\frac{(\d_c-\d_n)^2}{2(S_n-S_j)}
\right\}\, ,
\ee
where $A^2=\a\eps$ regularizes the integral over $dS_j$ when
$S_j\ra S_i^+$, and we want to extract the finite part as $A\ra 0$.
Observe also that here one must be careful not  to interchange the
limit ${\d_n\ra \d_c^-}$ (which comes from the  fact that the integral
over $d\d_n$ from $-\infty$ to $\d_c$ of
$\Pi^{(3,{\rm Lc})}_{\eps}$ is performed 
integrating by parts of
the derivative $\pa_n$ that appears in \eq{Pi3LcappB}) with the
integrals over $dS_i$ and $dS_j$. The integrals can be carried out
using the identity given in eq.~(A5) of paper~I, and we get
\be
I^{(d)}=-\frac{2}{\sqrt{2\pi}}
\langle\d_n^3\rangle\, \frac{1}{S_n^{3/2}}
\(\frac{\d_c}{A}+1\)\exp\left\{-\frac{(\d_c+A)^2}{2S_n}\right\}\, .
\ee
This has a part divergent as $1/A$, i.e. as $1/\sqrt{\eps}$, that will
combine with the similar divergences from the other terms, a part
finite as $A\ra 0$, plus terms ${\cal O}(A)$ that vanish in the continuum
limit. Extracting the finite part we get
\be\label{Id}
{\cal FP}[I^{(d)}]
=-2\, \frac{\langle\d_n^3\rangle}{\sqrt{2\pi}\, S_n^{3/2}}
\( 1-\frac{\d_c^2}{S_n}\)
\, e^{-\d_c^2/(2S_n)}\, .
\ee
Finally, the term $(e)$ can be computed with the by now usual trick of
replacing $\sum_{i,j,k=1}^{n-1}$ with $\pa^3/\pa \d_c^3$, and we get
\bees
\Pi^{(3,{\rm Le})}_{\eps}(\d_0;\d_n;S_n)&=& 
-\frac{\langle\d_n^3\rangle}{6}\, 
\sum_{i,j,k=1}^{n-1}
\int_{-\infty}^{\d_c} 
d\d_1\ldots d\d_{n-1}\pa_i\pa_j\pa_k W^{\rm gm}
=-\frac{\langle\d_n^3\rangle}{6}\, 
\frac{\pa^3}{\pa x^3_c}\Pi^{\rm gm}_{\eps}(\d_0;\d_c;S_n)\nn\\
&=& \frac{\langle\d_n^3\rangle}{6}\, 
\frac{4\sqrt{2}}{\sqrt{\pi}}\,  (2\d_c-\d_n)
\frac{1}{S_n^{5/2}}\[3-\frac{(2\d_c-\d_n)^2}{S_n}\]
\exp\left\{ -\frac{(2\d_c-\d_n)^2}{2S_n} \right\}
\, ,
\ees
and from this we find
\be\label{Ie}
I^{(e)}=\frac{4}{3}\, 
\frac{\langle\d_n^3\rangle}{\sqrt{2\pi}\, S_n^{3/2}}
\( 1-\frac{\d_c^2}{S_n}\)
\, e^{-\d_c^2/(2S_n)}\, .
\ee
Summing up \eqss{Ib}{Id}{Ie} we get back \eq{Pi3finiteapp}, as it should.

\end{document}